\documentclass[11pt]{article}
\usepackage{amssymb,eucal,amsmath} 
\usepackage{graphicx}
\usepackage{authblk}
\usepackage{cite}
\usepackage{calc}
\usepackage{accents}

\newcommand{\al}{\alpha}
\newcommand{\as}{\alpha_{\mathrm{s}}}   
\newcommand{\asb}{\bar{\al}_s}		
\newcommand{\bbar}{\bar{b}}         
\newcommand{\coll}{\mathrm{coll}}	
\newcommand{\de}{\partial}
\newcommand{\dbtilde}[1]{\tilde{\raisebox{0pt}[0.85\height]{$\tilde{#1}$}}}
\newcommand{\dif}{{\rm d}}
\newcommand{\dug}{\;\raisebox{0.07ex}{$:$}\hspace{-0.7ex}=}  
\newcommand{\dfn}{\,\raisebox{0.37pt}{:}\hspace{-3.2pt}=}

\newcommand{\eff}{\mathrm{eff}}
\newcommand{\ggf}{\CMcal{G}}		
\newcommand{\ggfb}{\mathrm{G}}		
\newcommand{\half}{\mbox{\small $\frac{1}{2}$}}
\renewcommand{\hom}{\frac{\om}{2}}        
\newcommand{\imf}{\phi}			
\newcommand{\imp}{\Longrightarrow}       
\newcommand{\krn}{\CMcal{K}}		
\newcommand{\lra}{\leftrightarrow}
\newcommand{\om}{\omega}
\newcommand{\ord}[1]{\mathcal{O}(#1)}   
\newcommand{\ordd}[1]{\mathcal{O}\left(#1\right)}   
\newcommand{\plq}{\mathcal{P}_{L}}		
\newcommand{\tr}{\mathrm{tr}}           
\newcommand{\ui}{{\rm i}}
\newcommand{\kt}{{\boldsymbol k}}       

\usepackage{color}
\usepackage{cleveref}
\usepackage[makeroom]{cancel}
\usepackage[normalem]{ulem}

\title{\bf Renormalization group improved  photon impact factors and  the high energy virtual photon scattering}
\author[1]{Dimitri Colferai}
\author[2]{Wanchen Li}
\author[2]{ Anna M. Sta\'sto}
\affil[1]{\small \it Dipartimento di Fisica, Universit\`a di Firenze and INFN Firenze
Via Sansone 1, 50019 Sesto Fiorentino, Italy}
\affil[2]{\small \it Department of Physics, Penn State University, University Park, PA 16802, USA}

\begin{document}
\maketitle

\begin{abstract}
We perform the renormalization group improved collinear resummation of the photon-gluon impact factors. 
We construct the resummed cross section for virtual photon-photon ($\gamma^*\gamma^*$) scattering which incorporates the impact factors and BFKL gluon Green's function up to the next-to-leading logarithmic accuracy in energy.
The impact factors include important kinematical effects which are responsible for the most singular poles in Mellin space at next-to-leading order. Further conditions on the resummed cross section are obtained by requiring the consistency with the collinear limits. Our analysis is consistent with previous impact factor calculations at NLO, apart from a new term proportional to $C_F$ that we find for the longitudinal polarization. Finally, we use the resummed cross section to compare with the LEP data on the $\gamma^*\gamma^*$ cross section and with previous calculations. The resummed result is lower than the leading logarithmic approximation but higher than the pure next-to-leading one, and is consistent with the experimental data.
\end{abstract}

\section{Introduction}

High energy particle accelerators, like the Large Hadron Collider (LHC),  opened up a new kinematic regime for particle interactions. Exploration of this regime is not only important for the phenomenological description of the scattering processes which occur at these colliders, but also for advancing our understanding of the theory of strong interactions: Quantum Chromodynamics (QCD). The high energy limit --- also called Regge limit --- in QCD is defined when the center-of-mass energy squared $s$ of the collision is much larger than other scales in the process, $s \gg -t  > \Lambda_{QCD}^2$, where $t$ is the momentum transfer. In the perturbative regime of small coupling $\alpha_s \ll 1$, the description of high energy processes has been developed over the decades, which is based on the high energy factorization (or $k_T$ factorization) \cite{Catani:1990eg,Catani:1990xk,Catani:1994sq} framework. The cross sections in this limit can be written in a factorized form with process-dependent impact factors and the universal gluon Green's function (GGF) responsible for the exchanges in the $t$-channel. It is the energy dependence of the latter that controls the high-energy behavior of the resulting cross section.  

The GGF is given by the solution to the Balitsky-Fadin-Kuraev-Lipatov (BFKL) \cite{Kuraev:1977fs,Balitsky:1978ic,Lipatov:1985uk} evolution equation, which resums the powers of $(\alpha_s \ln s/s_0)$, where $s_0$ is some reference energy scale. In the high energy limit, the logarithms of energy can be very large, and the terms $(\alpha_s \ln s/s_0)^n \sim 1$  even in the perturbative regime of small coupling. Therefore such terms need to be resummed, and this leads to the power growth of the gluon Green's function with the energy, that in turn translates to the power-like rise of the cross sections with energy. 
Due to the latter feature, this solution is traditionally referred to as the BFKL Pomeron.
The BFKL evolution equation is known at the leading logarithmic (LL) \cite{Kuraev:1977fs,Balitsky:1978ic} and next-to-leading logarithmic (NLL) accuracy in QCD \cite{Fadin:1998py,Ciafaloni:1998gs}.

As it turned out, the NLL corrections to the BFKL equation are large and negative, and may lead to instabilities, like oscillating cross section.  Thus, in order to stabilize the BFKL expansion, resummation methods were developed some time ago\ \cite{Salam:1998tj,Salam:1999cn,Altarelli:1999vw,Altarelli:2000mh,Altarelli:2001ji,Altarelli:2003hk,Altarelli:2008aj,Ciafaloni:1999au,Ciafaloni:1999yw,Ciafaloni:2003ek,Ciafaloni:2003rd,Ciafaloni:2003kd,Ciafaloni:2007gf,Thorne:2001nr,SabioVera:2005tiv}, and more recently applied to phenomenology  \cite{Bonvini:2016wki,Bonvini:2017ogt,Ball:2017otu}.   In the approach developed in \cite{Ciafaloni:2003ek,Ciafaloni:2003kd,Ciafaloni:2003rd,Ciafaloni:2007gf}, a renormalization group improved (RGI) small-$x$ evolution equation was constructed, which takes into account LL and NLL BFKL as well as the DGLAP splitting function at lowest order. The consistency of this formalism is based on the fact that the kernel of the evolution equation has the correct collinear limits, i.e., the limits of the strong ordering of the transverse momenta along the ladder of gluon emissions in the $t$-channel. To be precise, the requirement is that the collinear singularities are single logarithmic in transverse momenta, which in the Mellin space of the variable $\gamma$, the variable conjugated to the gluon transverse momentum, manifests itself as the occurrence of single poles for each power of $\as$.  The BFKL kernel in Mellin space has only single poles of type $\sim 1/\gamma,1/(1-\gamma) $  at leading logarithmic order, but at NLL order quadratic and cubic poles appear. The quadratic poles have been recognized as originating from the non-singular parts of the LO DGLAP splitting function which appears in the NLL BFKL kernel, as well as due to the running coupling. The cubic poles originate from the energy scales which become relevant at this order. It was demonstrated that the terms with the quadratic and cubic poles are the ones that are responsible for the major part of the NLL correction. 

In the following, by ``resummation'' we mean the ``collinear resummation'' of DGLAP terms, which coincides with the ``renormalization group improvement'' mentioned before.
In the Ciafaloni-Colferai-Salam-Stasto (CCSS) resummation scheme, the cubic poles at the NLL level (and the poles of order $2n+1$ at N$^n$LL level)
are resummed --- hence eliminated ---
by shifting the single poles in $\gamma$ in the LL kernel eigenvalue. The shift is proportional to the Mellin variable $\omega$ conjugated to the energy $s$.  This shift originates from the kinematical constraint \cite{Kwiecinski:1996td} imposed onto the integrals over the transverse momenta.  The quadratic poles originate from the non-singular part of the DGLAP splitting function, and again, can be resummed --- hence eliminated --- by taking them into account in the leading order kernel with an $\om$-dependent redefinition of the coefficients of the single poles. 

For a physical process that occurs at high energy, the gluon Green's function needs to be supplemented by the process-dependent impact factors, which also need to be evaluated at the appropriate order of perturbation theory.
The  NLO corrections have been calculated for the photon-gluon impact factor~\cite{Bartels:2002uz,Balitsky:2012bs}, Mueller-Navelet jet vertices \cite{Colferai:2010wu}, Mueller-Tang jets~\cite{Hentschinski:2014bra,
Hentschinski:2014esa,Colferai:2023hje}, and light vector mesons \cite{Ivanov:2004pp}. 
Numerous NLO calculations of impact factors have also been performed in the context of effective theory for high energy and density, the Color Glass Condensate, which includes parton saturation. Examples of the next-to-leading calculations in this framework include inclusive structure functions \cite{Beuf:2016wdz,Beuf:2017bpd}, also for massive quarks \cite{Beuf:2022ndu}, contributions to inclusive diffraction \cite{Beuf:2022kyp}, exclusive vector meson production 
\cite{Boussarie:2016bkq,Boussarie:2019ero,Mantysaari:2022kdm,Mantysaari:2022bsp} and for inclusive dijet \cite{Caucal:2022ulg} and photon+jet \cite{Roy:2019hwr}. The collinear resummation in the context of small $x$ evolution with saturation has been also explored, e.g.  \cite{Iancu:2015joa}.

An excellent process for studying the BFKL Pomeron  is $\gamma^*\gamma^*$ scattering, see e.g. \cite{Brodsky:1996sg,Brodsky:1997sd,Bartels:1996ke,Bartels:2000sk,Donnachie:1999kp,Donnachie:1999py, Kwiecinski:1999yx,Kwiecinski:2000zs}. The idea is to select the events in which the virtualities $Q_i^2$ of the two photons are comparable and large and the ``rapidity interval" $Y\equiv\log(s/Q_1 Q_2)$ between them is very large: $s\gg Q_1^2\sim Q_2^2\gg\Lambda_{QCD}^2$. In such kinematics, the DGLAP logarithms are suppressed and the process should be dominated by the BFKL evolution. This measurement was performed at LEP $e^+e^-$ collider~\cite{L3:2001uuv,OPAL:2001fqu}, through measurements of events with double tagged leptons.  Calculations were performed to describe this process within the BFKL formalism. 
In particular, it was shown that including the partial resummation in the gluon Green's function leads to an excellent description of the LEP data \cite{Kwiecinski:2000zs}.
Later on, once the NLO photon-gluon impact factor became available, the full NLL calculation of the photon-photon scattering was performed \cite{Chirilli:2014dcb,Ivanov:2014hpa}. The numerical difference between the LL and NLL case was found very large. In fact, the NLL calculation is not able to describe the  LEP data, particularly at the highest rapidity \cite{Ivanov:2014hpa}.

Given that the gluon Green's function required the collinear resummation, it is important to perform consistently the resummation of the impact factors. As in the case of the gluon Green's function it is important to take into account exact kinematics in the impact factors and analyze the structure of the poles due to these effects.  The exact kinematics was included in the $k_T$ factorization formula for the DIS structure functions \cite{Askew:1992tw,Askew:1993jk,Kwiecinski:1997ee}.  Based on this result, the photon-gluon impact factor with exact kinematics was computed in Mellin space~\cite{Bialas:2001ks}. Interestingly, these improved impact factors contained a shift of the poles in the $\gamma$ variable, analogously to what was observed in the Green's function with the kinematical constraint. This shift is proportional to $\om$, the Mellin variable conjugated to energy.

In this paper, we analyze the photon-photon scattering process at high energy and perform the resummation of the impact factors in addition to the resummation of the gluon Green's function. We construct the renormalization group improved high-energy factorization formula, where both impact factors and gluon Green's function are resummed, hence $\omega$-dependent. A first consistency condition is imposed to ensure the equivalence of the RGI impact factors and gluon Green's function upon expansion in $\omega$ with fixed order BFKL results up to NLL. 
The second consistency condition is imposed by analyzing the cross section in the collinear limit, i.e.,
assuming strong ordering of the virtualities of the photons, and consequently in the ladder of exchanged partons.
In that way the coefficients of the highest and next-to-highest $\gamma$-poles can be fixed both in the gluon Green's function and in the impact factors.  Since the two consistency conditions do not uniquely specify all the subleading poles when $\om\neq 0$, we consider several resummation schemes, which parametrize the ambiguity due to the unknown lower order poles.

As previously observed (see eg. \cite{Kwiecinski:2000zs,Ivanov:2014hpa}), we need to add to the BFKL cross section other contributions, in particular the one stemming from the quark box diagram --- both photons coupled to the same quark line ---, which is dominant for the lowest rapidities.
The results of our calculations are compared with the experimental data from LEP \cite{L3:2001uuv,OPAL:2001fqu} and an overall agreement is obtained within the theoretical and experimental uncertainties, with the resummed BFKL contribution representing the bulk of the cross section at high rapidities.

The paper is organized as follows. In sec.~\ref{s:greenfunction} we recall the renormalization group improved method for the gluon Green's function, and in particular  we discuss the $\omega$ shifts. In sec.~\ref{s:hef} the RGI factorization formula is introduced and consistency with high-energy factorization is discussed. We perform the collinear analysis of the transverse-transverse photon cross section in sec.~\ref{s:ritp} and  construct the resummed impact factor for transversely polarized photons. In sec.~\ref{s:rgilif} the analogous construction is carried out for the RGI impact factor in case of  longitudinally polarized photons. 
Numerical analysis is performed in
 sec.~\ref{s:numerical}, where  we  apply the  resummed impact factors to cross sections and compare the results with the experimental data from  LEP and with other theoretical descriptions. Finally, in sec.~\ref{s:conclusions} we state our conclusions. An appendix contains some formulae on the lowest order cross sections and structure functions.

\section{Renormalization group improved gluon Green's function\label{s:greenfunction}}
\label{sec:RGIgluon}

The collinear resummation for the evolution of gluon density at small-$x$ was developed in series of works using slightly different approaches \cite{Altarelli:1999vw,Altarelli:2000mh,Altarelli:2001ji,Altarelli:2003hk,Altarelli:2008aj}, \cite{Ciafaloni:1999au,Ciafaloni:1999yw,Ciafaloni:2003ek,Ciafaloni:2003kd,Ciafaloni:2003rd,Ciafaloni:2007gf}, \cite{Thorne:2001nr}, \cite{SabioVera:2005tiv}.  In this section we shall recap the essential ingredients of the resummation developed in \cite{Ciafaloni:1999au,Ciafaloni:1999yw,Ciafaloni:2003ek,Ciafaloni:2003kd,Ciafaloni:2003rd,Ciafaloni:2007gf} for the gluon Green's function. As a result, the renormalization group improved (RGI) small-$x$ evolution was constructed which contains both DGLAP and BFKL kernels and satisfies momentum sum rule. 

\begin{figure}[htp]
  \centering
  \includegraphics[width=0.3\linewidth]{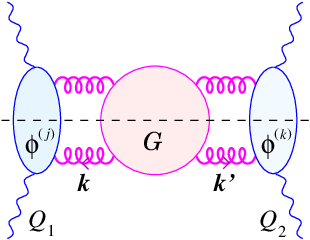}
  \caption{Diagramatic representation of the BFKL factorization formula for the process of high-energy scattering of two virtual photons.\label{f:hef}}
\end{figure}

We start from  the cross section $\sigma^{(jk)}$ for virtual photon
scattering at high-energy which can be written in a factorized form as the product of
process-dependent impact factors $\imf^{(j)}$ and the universal
(energy-dependent) gluon Green's function $\ggfb$, as depicted in \cref{f:hef}.
In momentum space, the BFKL factorization formula reads
\begin{equation}
  \sigma^{(jk)}(s,Q_1,Q_2) = \int\dif^2\kt\;\dif^2\kt' \;
    \imf^{(j)}(Q_1,\kt)\, \ggfb(s,\kt,\kt')\, \imf^{(k)}(Q_2,\kt') \;, \label{sigmaMomSp}
\end{equation}
where $j,k\in\{L,T\}$ denote the polarizations of the two photons, $q_1,q_2$ their momenta and $Q_i^2\equiv -q_i^2>0:i=1,2$ their virtualities.

The gluon Green's function $\ggfb(s,\kt,\kt_0) $, which depends on the transverse gluon momenta $\kt$ and $\kt_0$ and energy squared $s\equiv(q_1+q_2)^2$, satisfies the  evolution equation that can be written in the following form 
\begin{equation}
\frac{\partial }{\partial \log s} \ggfb(s,\kt,\kt_0) =  \int \dif^2  \kt' \; \krn( \kt,\kt') \, \ggfb(s,\kt',\kt_0) \; ,
    \label{eq:BFKL_Green_mom}
\end{equation}
where the function $\krn$ is the BFKL kernel which has the following perturbative expansion
\begin{equation}
\krn = \asb \krn_0 + \asb^2 \krn_1 + \dots \; .
    \label{eq:kernel_expansion}
\end{equation}
In the above equation, we introduced the rescaled strong coupling \(\asb = \frac{\alpha_s N_c}{\pi} \) where $N_c$ is the number of colors. In QCD the kernel is known at leading 
\cite{Kuraev:1977fs,Balitsky:1978ic} and next-to-leading order \cite{Fadin:1998py,Ciafaloni:1998gs}, in $\CMcal{N}=4$ super Yang-Mills theory up to next-to-next-to leading accuracy \cite{Velizhanin:2015xsa,Gromov:2015vua,caron2018high}. 
It is customary  to use the Mellin transform  to obtain the kernel eigenvalue
\begin{equation}
\asb \chi(\gamma) =  \int \dif\kt^{\prime 2} \; \bigg(\frac{\kt^{\prime2}}{\kt^2}\bigg)^{\gamma} \,  \krn(\kt,\kt') \, ,
    \label{eq:chi}
\end{equation}
with the corresponding perturbative expansion corresponding to \cref{eq:kernel_expansion}
\begin{equation}\label{asbchi}
     \chi(\gamma) =  \chi_0(\gamma) + \asb \chi_1(\gamma)+\dots \; .
\end{equation}
The leading  order kernel's eigenvalue reads
\begin{equation}\label{chi0}
  \chi_0(\gamma) =2\psi(1) - \psi(\gamma) - \psi(1-\gamma) \; ,
\end{equation}
where \(\psi(z) = \Gamma'(z)/\Gamma(z) \) is the polygamma function, and $\psi(1)=-\gamma_E$. The next-to-leading order eigenvalue is%
\footnote{This expression for $\chi_1$ holds for the scale-invariant part of the NLO kernel with symmetric energy scale (see \cref{sigmaBFKLandGGF}) and symmetric running coupling
$\as(|\kt|\,|\kt'|)$ in \cref{eq:kernel_expansion}.}
\begin{align}
	\chi_1(\gamma) &= -\frac{b}{2}
 \chi_0^2(\gamma)
 -\frac{1}{4}\chi_0''(\gamma) -\frac{1}{4}\left(\frac{\pi}{\sin \pi \gamma}\right)^2\frac{\cos \pi \gamma}{3(1-2\gamma)}\left[11+4\frac{T_R N_f}{N_c^3}+\frac{(1+2 T_R N_f/N_c^3)\gamma(1-\gamma)}{(1+2\gamma)(3-2\gamma)}\right] \nonumber \\
	&\quad +\left(\frac{67}{36}-\frac{\pi^2}{12}-\frac{5 T_R N_f}{9N_c}\right)\chi_0(\gamma) + \frac{3}{2}\xi(3)+\frac{\pi^2}{4\sin \pi \gamma } - \Phi(\gamma) \, ,
	\label{eq:nllorg}
\end{align}
where
\begin{equation}\label{bbar}
  b = \frac{11 N_c-4 T_R N_f}{12\pi} \equiv \frac{C_A}{\pi} \bbar     \, ,
\end{equation}
is the first beta-function coefficient, \(N_f\) the number of active quark flavours, $T_R=1/2$ and 
\begin{equation}
	\Phi(\gamma) = \sum_{n=0}^{\infty} (-1)^n\left[\frac{\psi(n+1+\gamma)-\psi(1)}{(n+\gamma)^2}+\frac{\psi(n+2-\gamma)-\psi(1)}{(n+1+\gamma)^2}\right] \, .
\end{equation}

The LO and NLO BFKL eigenvalues contain collinear and anticollinear poles, i.e. poles when \(\gamma \sim 0 \) and  \(\gamma \sim 1 \). These correspond to the strong ordering of the transverse momenta in the \( t\) channel, either $\kt^2 \gg \kt'^2$  or \(\kt^2 \ll \kt'^2\) respectively.

The LO and NLO eigenvalues have specific pole structures in \(\gamma\) variable. The LO eigenvalue has only single poles, i.e.
\begin{equation}\label{chi0poles}
  \chi_0(\gamma) \, \sim  \, \frac{1}{\gamma},\;\; \frac{1}{1-\gamma} \; .
\end{equation}
The NLO eigenvalue has a more complicated structure, since in addition to the single poles it also has double and triple collinear poles.  It is these higher order poles that are responsible for the fact that the NLO term is numerically large, and these terms need to be properly resummed. The double poles stem from two sources. The first one is  the running coupling
term
\begin{equation}
   - \frac{b}{2}\left[\chi_0^2(\gamma)
   \right] \, \sim  \, -\frac{b}{2\,\gamma^2} +
    (\gamma\to1-\gamma) \;,
   \label{eq:chi1_running_coupling}
\end{equation}
which contributes to the poles at $\gamma=0$ and 1. The second is the term
\begin{equation}
    -\frac{1}{4}\left(\frac{\pi}{\sin \pi \gamma}\right)^2\frac{\cos \pi \gamma}{3(1-2\gamma)}\left[11+
    4\frac{T_R N_f}{N_c^3}+
    \frac{(1+2 T_R N_f/N_c^3)\gamma(1-\gamma)}{(1+2\gamma)(3-2\gamma)}\right]
    \sim -\frac{11+4 T_R N_f/N_c^3}{12\;\gamma^2}+
    (\gamma\to1-\gamma) \; ,
    \label{eq:chi1_dglap_nonsing}
\end{equation}
which originates from the DGLAP anomalous dimension. To be precise, the coefficient of the double poles is just the non-singular part \(A_1(0)= -\frac{11+4 T_R N_f/N_c^3}{12}\) of the eigenvalue $\gamma_+(\om)$ of the LO DGLAP anomalous dimension matrix:
\begin{equation}
\gamma_+(\om) = P_{gg}(\om) +\frac{C_F}{C_A}P_{qg}(\om)+\ord{\om} = \frac{2 C_A}{\om}\left[1+\om A_1(\om)\right]  \; ,
\end{equation}
where \(\omega\) is the Mellin variable conjugated to the energy, \(C_A=N_c=3\) and $C_F=(N_c^2-1)/(2N_c)$
and
\begin{equation}\label{A1}
    A_1(\omega) = -\frac{1}{\omega+1}+\frac{1}{\omega+2}-\frac{1}{\omega+3}-\left[\psi(2+\omega)-\psi(1)\right] + \frac{11}{12} -\frac{T_R N_f}{3N_c^3} \, ,
\end{equation}
represents the non-singular (for $\om\to 0$) part of $\gamma_+$.

Finally, the triple collinear poles stem from the term
\begin{equation}
  -\frac{1}{4}\chi_0''(\gamma) \sim -\frac{1}{2} \frac{1}{\gamma^3}, \;\; -\frac{1}{2} \frac{1}{(1-\gamma)^3} \; .
    \label{eq:chi1_chidprime}
\end{equation}
The form of the term above depends on the scale choice for the kernel. 
Let us briefly recap the problem of energy scales \cite{Salam:1998tj} in the BFKL equation. 

Going back to the momentum representation of the BFKL equation~\eqref{eq:BFKL_Green_mom}, we can use the double Mellin transform to write the azimuthally averaged gluon Green's function as
\begin{equation} 
\ggfb( s, \kt, \kt_0 ) = \frac{1}{2\pi \kt^2}  \int\frac{\dif \omega}{2 \pi \ui}
  \bigg(\frac{s}{\kt\kt_0}\bigg)^{\om}  \int \frac{\dif\gamma}{2\pi \ui}\bigg( \frac{\kt^2}{\kt_0^2}\bigg)^{\gamma}  \ggfb(\om,\gamma) \, ,
  \label{eq:CG_gamma} 
\end{equation}
and the BFKL equation becomes
\begin{equation}\label{GomgammaEq}
    \omega\,\ggfb(\om,\gamma)  = 1+ \asb\, \chi(\gamma)\, \ggfb(\om,\gamma) \;.
\end{equation}
In \cref{eq:CG_gamma} we are adopting the symmetric energy scale $s_0=\kt\kt_0$.
However, the scale choice can also be asymmetric, like in the case of the Deep Inelastic Scattering, where the scales on the virtual photon and the proton side are in principle very different. In this case, the cross section is dominated by configurations with $\kt\gg\kt_0$ so that the proper evolution variable is $\kt^2/s$, corresponding to the asymmetric energy scale $s_0=\kt^2$.
The crucial observation is that such a change of energy scale in \cref{eq:CG_gamma} is equivalent to a shift of $\gamma$ by $\omega/2$:
\begin{equation} 
  \bigg(\frac{s}{\kt \kt_0}\bigg)^{\om}  \bigg( \frac{\kt^2}{\kt_0^2}\bigg)^{\gamma}=
  \bigg(\frac{s}{\kt^2}\bigg)^{\om} \bigg( \frac{\kt}{\kt_0}\bigg)^{\om}  \bigg( \frac{\kt^2}{\kt_0^2}\bigg)^{\gamma}=
  \bigg(\frac{s}{ \kt^2}\bigg)^{\om}  \bigg( \frac{\kt^2}{\kt_0^2}\bigg)^{\gamma+\om/2} \;.
  \label{eq:different_scales} 
\end{equation}
The opposite shift is obtained if $s_0=\kt_0^2$.
Due to that fact, the gluon Green's function and thus the kernel in Mellin space gets $\omega$ dependence; the latter can be written in the form~\cite{Ciafaloni:1998iv}
\begin{equation}
   \chi_0(\omega,\gamma)= 2\psi(1) - \psi(\gamma + \frac{\omega}{2}) - \psi(1-\gamma+ \frac{\omega}{2}) \; ,
   \label{eq:chi_om_sym}
\end{equation}
for the symmetric scale choice.

Expanding this kernel in $\omega$ and using the solution at lowest order $\omega=\asb \chi_0(\gamma)$, one obtains
for the NLO contribution 
\begin{equation}
-\frac{1}{2}\omega\,  \psi'(\gamma) -\frac{1}{2}\omega \, \psi'(1-\gamma) \; \simeq \; -\frac{1}{2} \frac{\asb}{\gamma^3}  -\frac{1}{2} \frac{\asb}{(1-\gamma)^3}  \;.
\label{eq:triple_poles}
\end{equation}
These terms exactly correspond to the triple collinear poles present in the NLO kernel, see \cref{eq:chi1_chidprime}.
In other words, the cubic poles in the NLO kernel can be discarded, since their contribution is taken into account by the $\om$-shift in the LO kernel, as in \cref{eq:chi_om_sym}.

Next, the collinear term with the non-singular DGLAP splitting function was included in the form~\cite{Ciafaloni:2003rd}
\begin{equation}\label{eq:chic}
   \chi_c^{\om}(\gamma)= \omega A_1(\omega)\left(\frac{1}{\gamma+\frac{\omega}{2}} + \frac{1}{1-\gamma+\frac{\omega}{2}} \right) \; ,
\end{equation}
which, when expanded in $\omega$ and retaining the first power in $\asb$, gives
\begin{equation}\label{eq:double_poles}
    \omega A_1(\omega)\left(\frac{1}{\gamma}+\frac{1}{1-\gamma}\right) \simeq \asb A_1(0) \left( \frac{1}{\gamma^2}+\frac{1}{(1-\gamma)^2} \right) \; ,
\end{equation}
thus reproducing \cref{eq:chi1_dglap_nonsing}. These terms are then subtracted from the NLO kernel again.

Actually, the $\om$-shift predicted by the collinear analysis with upper and lower energy-scale, leading to \cref{eq:chi_om_sym}, allows us to predict the spurious poles
\footnote{It has been verified \cite{Deak:2019wms} that the $\om$-shift correctly reproduces the highest order poles $\sim 1/\gamma^5$ at the NNLO BFKL in the supersymmetric theory \cite{Gromov:2015vua,Velizhanin:2015xsa,caron2018high}.}
of the higher order BFKL kernels:
\begin{equation}
  \chi_n(\gamma) \sim \frac1{\gamma^{1+2n}} \; ,
\end{equation}
which are more and more singular as the order increases, while only poles of order $1+n$ are expected from the collinear QCD dynamics.
This can be roughly understood because the $\om$-shift transforms a LO pole into a series of spurious poles to all orders, e.g.,
\begin{equation}
  \frac1{\gamma+\om/2 } \sim \frac{1}{\gamma+\asb\chi_0/2}
  \sim\frac{1}{\gamma(1+\asb/2\gamma^2)} \sim
  \sum_{n=0}^\infty \frac{(-\asb)^n}{2^n\,\gamma^{1+2n}} \;.
\end{equation}

The occurrence of ``spurious" high-order poles in the BFKL approach is responsible for the bad convergence of the BFKL expansion and the instabilities of its phenomenological predictions. Therefore, it is compelling to resum such spurious poles by means of the RGI formulation.

In conclusion, the resummed kernel in the CCSS formalism was constructed by taking at LO the sum of the $\om$-shifted kernels in \cref{eq:chi_om_sym,eq:chic}
and subtracting the triple and double $\gamma$-poles (\cref{eq:triple_poles,eq:double_poles}) from the NLO kernel. In this way the NLO resummed kernel has only simple poles, hence it is much less singular in the collinear limits $\gamma\to 0,1$ and provides more stable and reliable phenomenological results.

\section{High energy factorisation\label{s:hef}}

In this section we recall and compare the factorization formulae in the pure BFKL formalism and in the RGI approach, with the aim of deriving the compatibility conditions among the respective impact factors and Green functions, thus setting the stage for the computation of the RGI impact factors.

\subsection{BFKL vs RGI factorization formula\label{s:bvrff}}

The high-energy factorization formula~\eqref{sigmaMomSp} for $\gamma^* \gamma^*$ scattering can be rewritten in a more convenient form as a double Mellin representation with respect to transverse momenta (or virtualities)
[cfr.~\cref{eq:chi}] and to energy [cfr.~\cref{eq:CG_gamma,GomgammaEq}]:
\begin{subequations}\label{sigmaBFKLandGGF}
  \begin{align}
  \sigma^{(jk)}(s,Q_1,Q_2) &= \frac1{2\pi Q_1 Q_2}
    \int\frac{\dif \omega}{2\pi\ui} \left(\frac{s}{s_0(p)}\right)^\om
    \int\frac{\dif \gamma}{2\pi\ui} \left(\frac{Q_1^2}{Q_2^2}\right)^{\gamma-\half}
    \imf^{(j)}(\gamma;p)\, \ggfb(\om,\gamma;p)\, \imf^{(k)}(1-\gamma;-p) \, ,\label{sigmaBFKL} \\
    \ggfb(\om,\gamma;p) &= \frac1{\om-\asb\chi(\gamma;p)}\;. \label{BFKLggf}
  \end{align}
\end{subequations}
Here we introduced the notation $s_0(p) = Q_1^{1+p} Q_2^{1-p}$ for the energy-scale. By varying the
parameter $p$ we can switch from symmetric scale $s_0=Q_1 Q_2$ ($p=0$), to
``upper'' scale $s_0=Q_1^2$ ($p=1$) or to ``lower'' scale $s_0=Q_2^2$ ($p=-1$).

In \cref{sigmaBFKLandGGF} both impact factors $\imf$ and eigenvalue
function $\chi$ are perturbative objects that admit a series expansion in $\as$, as in \cref{asbchi};
from next-to-leading order on, they depend on the choice of the energy scale:
\begin{align}
  \imf^{(j)}(\gamma;p) &= \imf^{(j)}_0(\gamma) + \asb \imf^{(j)}_1(\gamma;p) +
  \ord{\asb^2}\, , \label{PhiBFKL} \\
  \chi(\gamma;p) &= \chi_0(\gamma) + \asb \chi_1(\gamma;p) +
  \ord{\asb^2} \;. \label{chiBFKL}
\end{align}

On the other hand, the renormalization-group improved (RGI) high-energy factorization for scattering reads
\begin{subequations}\label{sigmaRGIandGGF}
  \begin{align}
    \sigma^{(jk)}(s,Q_1,Q_2) &= \frac1{2\pi Q_1 Q_2}
    \int\frac{\dif \omega}{2\pi\ui} \left(\frac{s}{s_0(p)}\right)^\om
    \int\frac{\dif \gamma}{2\pi\ui} \left(\frac{Q_1^2}{Q_2^2}\right)^{\gamma-\half}
    \Phi^{(j)}(\om,\gamma;p) \,\ggf(\om,\gamma;p)\,\Phi^{(k)}(\om,1-\gamma;-p) \label{sigmaRGI} \\
    \ggf(\om,\gamma;p) &= \frac1{\om-\asb X(\om,\gamma;p)} \;.
    \label{rgiGGF} 
  \end{align}
\end{subequations}
Here, we introduce the new notation $X(\om,\gamma)$ for the kernel in Mellin space appearing in RGI factorization~\eqref{rgiGGF} to clearly distinguish it from the BFKL kernel $\chi(\gamma)$ present in the standard high-energy factorization, \cref{BFKLggf}.
At variance with the usual BFKL expansion, both impact factors $\Phi$ and
eigenvalue function $X$ are $\om$-dependent, for a twofold purpose:
{\it (i)} to fully agree with the known collinear
behaviour at least in the leading-logarithmic $\log(Q_1/Q_2)$ approximation;
{\it (ii)} to resum into a smoother behaviour subleading contributions which are
singular in some region of the complex $\gamma$-plane. Actually, the two issues
are strictly related, as explained in~\cite{Ciafaloni:2003rd} and in sec.~\ref{s:greenfunction}.

Following the argument leading to \cref{eq:different_scales}, a change of the energy scale $s_0$, i.e.\ a change in $p$, leaves the cross
section~\eqref{sigmaRGI} invariant provided
\begin{align}
  \Phi^{(j)}(\om,\gamma;p) = \Phi^{(j)}\Big(\om,\gamma-\frac{\om}{2} p;0\Big) \;,\qquad
  X(\om,\gamma;p) = X\Big(\om,\gamma-\frac{\om}{2} p;0\Big)  \;. \label{scaleChange}
\end{align}
The corresponding scale change entails more complicated changes in the BFKL
impact factors $\imf$ and eigenvalue function $\chi$ of \cref{sigmaBFKLandGGF}.

Given some choice for the energy scale, the equivalence between the two
factorization formulas~\eqref{sigmaBFKL} and~\eqref{sigmaRGI} is obtained by
evaluating the $\om$-integrals and requiring the remaining $\gamma$-integrand to
be the same function (up to terms yielding contributions suppressed by powers of $s$). In \cref{sigmaBFKL} the $\om$ integration is trivial:
being $s_0 < s$ one can close the $\omega$-integration path to the left and pick
up the simple pole at $\om = \asb\chi(\gamma)$, obtaining
\begin{align}
  \sigma^{(jk)}(s,Q_1,Q_2) &= \frac1{2\pi Q_1 Q_2}
  \int\frac{\dif \gamma}{2\pi\ui} \left(\frac{s}{s_0}\right)^{\asb\chi(\gamma)}
  \left(\frac{Q_1^2}{Q_2^2}\right)^{\gamma-\half}
  \imf^{(j)}(\gamma)\, \imf^{(k)}(1-\gamma) \; .
  \label{sigmaBFKL2}
\end{align}

In \cref{sigmaRGI} there can be many $\omega$-poles. The
position of the rightmost pole --- which provides the leading high-energy behaviour of the cross section --- is determined by the implicit equation
\begin{equation}\label{omeff}
 \om=\asb X(\om,\gamma) \equiv
 \om^\eff(\gamma,\asb) \equiv
 \asb\chi^{\eff}(\gamma,\asb)\;,
\end{equation}
where the last expressions $\om^\eff=\asb\chi^{\eff}$ represent such solution as function of $\gamma$ and $\asb$.
Then the  $\omega$-integral singles out the residue at      such pole
\begin{equation}
  \mathrm{Res}_{\om=\om^\eff}[\om-\asb X(\om,\gamma)]^{-1}
  = [1-\asb\de_\om X(\om^\eff,\gamma)]^{-1} \;,
\end{equation}
yielding
\begin{align}
  \sigma^{jk}(s,Q_1,Q_2) &= \frac1{2\pi Q_1 Q_2}
  \int\frac{\dif \gamma}{2\pi\ui} \left(\frac{s}{s_0}\right)^{\asb X(\om^\eff,\gamma)}
  \left(\frac{Q_1^2}{Q_2^2}\right)^{\gamma-\half}
  \frac{\Phi^{(j)}(\om^\eff,\gamma)\, \Phi^{(k)}(\om^\eff,1-\gamma)}{
   1-\asb\de_\om X(\om^\eff,\gamma)}
  +\cdots \;,
  \label{sigmaRGI2}
\end{align}
where the dots indicate terms suppressed by powers of $s$.
Therefore, for any choice of energy scale,
\begin{subequations}\label{RGIrel}
  \begin{align}
  \chi(\gamma) &= X(\om^\eff,\gamma) \label{chirel} \\
  \imf^{(j)}(\gamma) \imf^{(k)}(1-\gamma) &=
  \frac{\Phi^{(j)}(\om^\eff,\gamma)\Phi^{(k)}(\om^\eff,1-\gamma)}{1-\asb\de_\om X(\om^\eff,\gamma)} \;. \label{IFrel}
\end{align}
\end{subequations}
By expanding \cref{RGIrel} in $\asb$ as in \cref{PhiBFKL,chiBFKL}, we obtain the following equations relating the RGI eigenvalue and impact factors (and their derivatives) at $\om=0$ with the BFKL ones: 
\begin{align}
  \om^\eff &= \asb \chi_0(\gamma) + \ord{\asb^2} \label{omeffexpn} \\
  \chi_0(\gamma) &= X_0(0,\gamma) \label{chi0expn} \\
  \chi_1(\gamma) &= X_1(0,\gamma) + \chi_0(\gamma) \partial_\om X_0(0,\gamma) \label{chi1expn} \\
  \imf_0^{(j)}(\gamma)\imf_0^{(k)}(1-\gamma) &= \Phi_0^{(j)}(0,\gamma)\Phi_0^{(k)}(0,1-\gamma) \label{F0expn} \\
  \imf_0^{(j)}(\gamma)\imf_1^{(k)}(1-\gamma)+\imf_1^{(j)}(\gamma)\imf_0^{(k)}(1-\gamma) &=
  \Phi_0^{(j)}(0,\gamma)\left[
    \Phi_1^{(k)}(0,1-\gamma)+\chi_0(1-\gamma)\de_\om \Phi_0^{(k)}(0,1-\gamma)\right]
  \nonumber \\
  &\quad+ \left[\Phi_1^{(j)}(0,\gamma)+\chi_0(\gamma)\de_\om \Phi_0^{(j)}(0,\gamma)
    \right]\Phi_0^{(k)}(0,1-\gamma) \nonumber \\
  &\quad+\Phi_0^{(j)}(0,\gamma)\Phi_0^{(k)}(0,1-\gamma)\de_\om X_0(0,\gamma) \;. \label{F1expn}
\end{align}
These equations form the first consistency condition of the RGI framework with the BFKL framework. 
\Cref{omeffexpn,chi0expn,chi1expn} are well known
from the first studies on RGI BFKL~\cite{Ciafaloni:2003rd}. \Cref{F0expn} implies that
$\imf_0^{(j)}(\gamma)=\Phi_0^{(j)}(0,\gamma)$ for any polarization $j$.
In particular, \cref{sigmaBFKL,sigmaRGI} imply the
following normalization for the LO impact factors, compared to those
of refs.~\cite{Catani:1994sq,Ivanov:2014hpa,Bialas:2001ks}:
\begin{align}
  \imf_0^{(j)}(\gamma) &= \frac{2\pi\sqrt{2(N_c^2-1)}\al}{N_f}
  \left(\sum_q e_q^2\right) \;
  \gamma\, h_j(\gamma)
  && (h_T=\frac{h_2}{\gamma}-h_L)
  &&\text{ref~\cite{Catani:1994sq} Catani et al.} \label{cfrCatani}\\
  &= \frac{T_R \sqrt{2(N_c^2-1)}}{2} F_j(\nu) 
  &&(\gamma=\half+\ui\nu)
  &&\text{ref~\cite{Ivanov:2014hpa} Ivanov et al.} \label{cfrIvanov}\\
  &= \frac{T_R\sqrt{2(N_c^2-1)}}{\pi}
  \left(\sum_q e_q^2\right) \; S_j(N=0,\gamma)
  && (N=\om)
  &&\text{ref~\cite{Bialas:2001ks} Bia\l as et al.} \;, \label{cfrBialas}
\end{align}
where $\sum_q$ denotes the sum over quark flavours and $e_q$ is the electric charge of quark $q$ in units of the positron charge.
In those papers, the expressions are often given for $N_c=3$ and $T_R = 1/2$,
but it is better to keep track of such colour structure for the
comparison with the subsequent collinear analysis.
Explicitly, the LO impact factors read
\begin{subequations}\label{phi0}
\begin{align}
  \imf_0^{(T)}(\gamma) &= \al \as \Big(\sum_q e_q^2\Big) T_R 
  \sqrt{2(N_c^2-1)}\,\frac{\pi}{2}\;
  \frac{(1+\gamma)(2-\gamma)\Gamma^2(\gamma)\Gamma^2(1-\gamma)}{
  (3-2\gamma)\Gamma(3/2+\gamma)\Gamma(3/2-\gamma)} \, ,\label{phi0T} \\
    \imf_0^{(L)}(\gamma) &= \al \as \Big(\sum_q e_q^2\Big) T_R 
  \sqrt{2(N_c^2-1)}\,\pi\;
  \frac{\Gamma(1+\gamma)\Gamma(2-\gamma)\Gamma(\gamma)\Gamma(1-\gamma)}{
  (3-2\gamma)\Gamma(3/2+\gamma)\Gamma(3/2-\gamma)} \label{phi0L} \;,
\end{align}
\end{subequations}
where $\alpha$ is the electromagnetic coupling.
It is apparent from the $\Gamma$ functions in the numerators that both LO impact factors have poles at $\gamma=0$ and $\gamma=1$, similarly to the eigenvalue functions $\chi_0$ and $\chi_1$ in \cref{chi0poles,eq:chi1_chidprime}. This is due to 
QCD dynamics which, in the collinear limit $Q_1\gg Q_2$ ($Q_1\ll Q_2$), generates logarithmic term $\sim\log^n(Q_1^2/Q_2^2):n\geq 0$, corresponding to poles of order $n+1$ at $\gamma=0$ ($\gamma=1$) in Mellin space. More precisely, the RGI impact factors and eigenvalue function have poles whose order increases as the perturbative order:
\begin{align}\label{PhiPoles}
  \Phi^{(T)}_n(\om,\gamma;1) &\sim \frac1{\gamma^{2+n}} \;,&
  \Phi^{(L)}_n(\om,\gamma;1) &\sim \frac1{\gamma^{1+n}} \;,&
  X_n(\om,\gamma;1) &\sim \frac1{\gamma^{1+n}} \;,
\end{align}
as will be evident from the collinear analysis in the next section. On the other hand, the corresponding BFKL quantities at symmetric scale $s_0=Q_1 Q_2$ have poles that increase twice as much: 
\begin{align}\label{imfPoles}
  \imf^{(T)}_n(\gamma;0) &\sim \frac1{\gamma^{2+2n}} \;,&
  \imf^{(L)}_n(\gamma;0) &\sim \frac1{\gamma^{1+2n}} \;,&
  \chi_n  (\gamma;0) &\sim \frac1{\gamma^{1+2n}} \;.
\end{align}
This has been already observed at leading and next-to-leading order for the eigenvalue functions $\chi_0$ and $\chi_1$ in sec.~\ref{s:greenfunction} and in particular in \cref{eq:chi1_chidprime}. The collinear poles of the NLO impact factors can be derived from the expressions computed in \cite{Ivanov:2014hpa}:
\begin{align}
\frac{\imf^{(T)}_1(\gamma;0)}{\imf^{(T)}_0(\gamma)} & = \frac{\chi_0(\gamma)}{2}\ln{\frac{s_0}{Q^2}} + 
    \bbar\ln{\frac{\mu_R^2}{Q^2}} \nonumber \\
    & + \frac{3C_F}{4N_C} - \frac{5}{9}\frac{T_R N_f}{N_C}+\frac{\pi^2}{4}+\frac{85}{36}-\frac{\pi^2}{\sin^2(\pi \gamma)} - \frac{1}{\gamma\left(\gamma-1 \right)} + \frac{3\chi_0(\gamma)}{2\left(\gamma+1 \right)\left(2-\gamma \right)} \nonumber\\
    & + \frac{1}{4(1-\gamma)} - \frac{1}{4\gamma} - \frac{7}{36(1+\gamma)} + \frac{5}{3(1+\gamma)^2} -\frac{25}{36(\gamma -2)} \nonumber\\
    & + \frac{1}{2}\chi_0(\gamma)\left[\psi\left(1-\gamma\right) +2\psi\left(2-\gamma\right) -2\psi\left(4-2\gamma\right) -\psi\left(2+\gamma\right) \right],
    \label{eq: Ivanov FT1}\\
    \frac{\imf^{(L)}_1(\gamma;0)}{\imf^{(L)}_0(\gamma)}& = \frac{\chi_0(\gamma)}{2}\ln{\frac{s_0}{Q^2}} + 
    \bbar\ln{\frac{\mu_R^2}{Q^2}} \nonumber \\
    & + \frac{3C_F}{4N_C} - \frac{5}{9}\frac{T_R N_f}{N_C}+\frac{\pi^2}{4}+\frac{85}{36}-\frac{\pi^2}{\sin^2(\pi \gamma)} - \frac{1-4\gamma}{2\gamma^2(\gamma^2 -1)}  + \frac{1}{1-\gamma^2}\chi_0(\gamma) \nonumber\\
    & + \frac{1}{2}\chi_0(\gamma)\left[\psi\left(1-\gamma\right) +2\psi\left(2-\gamma\right) -2\psi\left(4-2\gamma\right) -\psi\left(2+\gamma\right) \right],
    \label{eq: Ivanov FL1}
\end{align}
where $\bbar$ is defined in \cref{bbar} and $\mu_R$ is the renormalization scale.
The origin of the higher order poles in impact factors is the same as that of poles in the BFKL kernel, as explained in sec.~\ref{sec:RGIgluon}. Such spurious poles can be resummed using the $\om$-shift of poles suggested by the RGI procedure.

\Cref{F1expn} will be used to determine the RGI impact factors at
NLO. For this purpose, we need to know the $\om$-dependence of the LO
eigenvalue and impact factors. All that will be the subject of the next
section.

\section{RGI impact factor for transverse photons\label{s:ritp}}

\subsection{Lowest order $\boldsymbol{TT}$ cross section in the collinear limits\label{s:cscl}}

\begin{figure}[hbp]
  \centering
  \includegraphics[width=0.3\linewidth]{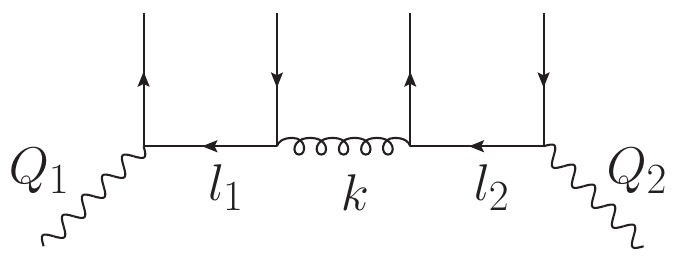}
  \caption{Diagramatics of collinear limit at lowest order in the BFKL factorization formula.\label{f:colChain}}
\end{figure}

Further information for the $\gamma^*\gamma^*$ cross section, somehow
complementary to the multi-Regge kinematics, can be inferred by analyzing the
collinear limit, i.e., by considering two photons with very different
virtualities, say $Q_1\gg Q_2$. This situation is well described by effective
ladder diagrams, like the one depicted in \cref{f:colChain}, where the
intermediate propagators are strongly ordered in virtuality (decreasing from
left to right). At each QCD vertex, the strong coupling is evaluated at a scale
given by the largest virtuality of the connected propagators, while a splitting
function $P_{ba}(z_b/z_a)$ describes the fragmentation of the parent parton $a$
(to the right) into a child parton $b$ (to the left) and an emitted on-shell parton
(vertical line). The integrals over the ordered longitudinal momentum fractions
are convolutions, which can be diagonalized by a Mellin transform in
the Bjorken variable $1/x_{Bj}=s/Q_1^2=s/s_0(p=1)$:
\begin{align}
  \sigma^{(jk)}(s,Q_1,Q_2) &= \frac{1}{2\pi Q_1 Q_2} \int\frac{\dif\om}{2\pi\ui}
  \left(\frac{s}{Q_1^{1+p}Q_2^{1-p}}\right)^\om\;
  \tilde{\sigma}^{(jk)}(\om,Q_1,Q_2;p) \;.\label{sigmaColTT}
\end{align}
which is exactly the structure of \cref{sigmaBFKL,sigmaRGI}.

The collinear integrand $\tilde{\sigma}^{(TT)}$ for two transverse photons at
$\ord{\al^2 \as^2}$ --- corresponding to the four-rungs LO BFKL diagram --- is
given by (cfr.~\cref{s:locs})
\begin{align}
    \tilde{\sigma}^{(TT)}(\om,Q_1,Q_2;1) &= 
  (2\pi)^3\al \Big(2\sum_{q\in A} e_q^2\Big) \times \nonumber\\
  & \int_{Q_2^2}^{Q_1^2}\frac{\dif l_1^2}{l_1^2} \frac{\as(l_1^2)}{2\pi} P_{qg}(\om)
  \int_{Q_2^2}^{l_1^2}\frac{\dif k^2}{k^2} \frac{\as(k^2)}{2\pi} P_{gq}(\om)
  \int_{Q_2^2}^{k^2}\frac{\dif l_2^2}{l_2^2} \frac{\al}{2\pi}
  \Big(2\sum_{q\in B} e_{q}^2\Big) P_{q\gamma}(\om) \;. \label{ITTQ}
\end{align}
where $l_1$, $k$ and $l_2$ are the momenta of the $t$-channel quark, gluon and
quark respectively, as depicted from left to right in \cref{f:colChain}, $A$ and $B$ denote the sets of active quarks of momenta $l_1$ and $l_2$ respectively,
while $P_{ab}(\om)$ denote Mellin moments of the one-loop splitting functions.
The running coupling at scale $|k|$ is defined in terms of the renormalized
coupling $\as$ at the renormalization scale $\mu_R$:
\begin{align}
  \as(k^2) &\dug
  \frac{\as(\mu_R^2)}{1+\as(\mu_R^2)b\ln\frac{k^2}{\mu_R^2}}
  \simeq \as(\mu_R^2)\left(1-\as(\mu_R^2)b\ln\frac{k^2}{\mu_R^2}+\cdots\right) \, ,
  \label{rc}
\end{align}
where $b$ is defined in \cref{bbar}. 
Substituting the above expansion for $\as(l_1^2)$ and $\as(k^2)$ in \cref{ITTQ} and switching to logarithmic variables
$L_i\dfn \ln\frac{Q_i^2}{\mu_R^2}$, $\lambda_i\dfn\ln\frac{l_i^2}{\mu_R^2}$,
$\lambda_k\dfn\ln\frac{k^2}{\mu_R^2}$, we obtain
\begin{align}
  \tilde{\sigma}^{(TT)}
  &= (2\pi)^3 \frac{\al^2}{2\pi} \left(\frac{\as(\mu_R^2)}{2\pi}\right)^2
  \Big(2\sum_{q\in A} e_q^2\Big) \Big(2\sum_{q\in B} e_{q}^2\Big)
   P_{qg}(\om) P_{gq}(\om) P_{q\gamma}(\om) \nonumber \\
  &\quad\times \int_{L_2}^{L_1}\dif\lambda_1 \int_{L_2}^{\lambda_1}\dif\lambda_k
  \int_{L_2}^{\lambda_k}\dif\lambda_2\;
  \left[1-\as(\mu_R^2)b(\lambda_1+\lambda_k)+\ord{\as^2}\right] \;.
  \label{nestInt}
\end{align}
The nested integral in the second line of \cref{nestInt} yields
\begin{align*}
  \iiint &= \frac{(L_1-L_2)^3}{3!} \left[1 - \as(\mu_R^2)b (L_1+L_2)\right]
  - \as(\mu_R^2)b \frac{(L_1-L_2)^4}{4!} +\ord{\as^2} \;.
\end{align*}
By including the overall factor $\as^2(\mu_R^2)$ written in the first line of \cref{nestInt} and noting that
\begin{align*}
  \as^2(\mu_R^2) \left[1 - \as(\mu_R^2)b (L_1+L_2)\right]
  = \as^2(Q_1 Q_2) + \ord{\as^4} \;,
\end{align*}
we get
\begin{align}\label{rcCorrk}
  \as^2(\mu_R^2) \iiint &\simeq
  \as^2(Q_1 Q_2)\left[ \frac{(L_1-L_2)^3}{3!}-\as b \frac{(L_1-L_2)^4}{4!}
    + \ord{\as^4}\right]\;, \\
   &\simeq
  \as^2(Q_1 Q_2)\left[ \frac{1}{3!}\log^3\frac{Q_1^2}{Q_2^2}
    -\as b \frac{1}{4!}\log^4\frac{Q_1^2}{Q_2^2}
    + \ord{\as^4}\right]\;.
\end{align}
By Mellin transforming in $Q_1^2/Q_2^2$ the terms in square brackets%
\footnote{Recall that $(L_1-L_2)^n=\ln^n\frac{Q_1^2}{Q_2^2}$ becomes $\frac{n!}{\gamma^{n+1}}$ under Mellin transform.}
(while keeping the strong couplings as factors outside the Mellin transform),
we obtain the corresponding expression in $\gamma$-space: 
\begin{align}\label{rcCorr}
  \as^2(Q_1 Q_2)\,\frac1{\gamma^4} \left[1 -\as b \frac1{\gamma}\right] \;.
\end{align}
The first term $\ord{\as^2/\gamma^4}$ could have been obtained by using a fixed
coupling constant in \cref{ITTQ}. The introduction of the running coupling
is responsible for the second ($b$-dependent) term $\ord{\as^3/\gamma^5}$,
which will be important in the analysis of the NLO impact factors.%
\footnote{If one chooses a different scale for the running coupling, the
coefficient of the $b$-dependent term in \cref{rcCorr}
would change accordingly.}

Finally, by restoring all the factors of \cref{nestInt}, we obtain the
Mellin transform of $\tilde{\sigma}^{(TT)}$ of \cref{ITTQ}
(with respect to the variable $Q_1^2/Q_2^2$), expanded at order $\as^2$,
which is nothing but the integrand of the RGI factorization
formula~\eqref{sigmaRGI} in the collinear limit $\gamma\to0$:
\begin{align}
  \dbtilde{\sigma}_0^{(TT)}(\om,\gamma;1)\big|^{\coll}
  &= \Phi^{(T)}_0\,\ggf_0\,\Phi^{(T)}_0\big|^{\coll}_{p=1} \nonumber\\
  &= (2\pi)^3\al \Big(2\sum_{q\in A} e_q^2\Big) \frac1{\gamma}
  \;\cdot\; \frac{\as}{2\pi} \frac{P_{qg}(\om)}{\gamma}
  \;\cdot\; \frac{\as}{2\pi} \frac{P_{gq}(\om)}{\gamma} \;\cdot\;
  \frac{\al}{2\pi} \Big(2\sum_{q\in B} e_{q}^2\Big) \frac{P_{q\gamma}(\om)}{\gamma} \;.
  \label{colChainLO}
\end{align}
Some remarks are in order:
\begin{itemize}
\item[{\it (i)}] Since the collinear analysis of the cross section based on the DGLAP
  chain singles out the leading logarithmic behaviour in the ratio $Q_1/Q_2$,
  \cref{colChainLO} provides just the leading $\gamma$-pole structure of the
  RGI integrand in the neighborhood of $\gamma=0$.
\item[{\it (ii)}] Such pole structure correspond to $p=1$, i.e., energy scale
  $s_0=Q_1^2$. Adopting the symmetric energy scale $s_0=Q_1Q_2$ ($p=0$),
  according to \cref{scaleChange} the pole at $\gamma=0$ is shifted at
  $\gamma=-\om/2$, while keeping the same coefficient.
\item[{\it (iii)}] In the opposite (anti)collinear limit $Q_1\ll Q_2$, one
  obtains the same result of \cref{colChainLO}, provided one replaces
  $\gamma\to 1-\gamma$ and $p\to -1$, i.e., $s_0=Q_2^2$. At symmetric energy
  scale $s_0=Q_1 Q_2$, the pole at $\gamma=1$ is shifted at $\gamma=1+\om/2$. If $p=1$,
  i.e., $s_0=Q_1^2$, the pole at $\gamma=1$ is shifted at $\gamma=1+\om$.
\item[{\it(iv)}] The two sums with electric charges are over quark flavours
  ($q\in\{u,d,\dots\}$) and a factor of 2 in front of each sum takes into account
  quark+antiquark contributions.
\end{itemize}

Therefore, with energy scale $s_0=Q_1^2$ ($p=1$) and including both
collinear and anticollinear contributions, the pole structure at LO of the RGI
improved cross section reads
\begin{align}
  \dbtilde{\sigma}_0^{(TT)}(\om,\gamma;1)\big|^{2\times\coll}
  &= (2\pi)^3\al \Big(2\sum_{q} e_q^2\Big)\frac1{\gamma}
  \cdot \frac{\as}{2\pi} \frac{P_{qg}(\om)}{\gamma}
  \cdot \frac{\as}{2\pi} \frac{P_{gq}(\om)}{\gamma}
  \cdot \frac{\al}{2\pi} \Big(2\sum_{q} e_{q}^2\Big)
  \frac{P_{q\gamma}(\om)}{\gamma} \nonumber \\
  &\qquad+ \Big(\gamma\to 1+\om-\gamma\Big) \;. \label{TGT0}
\end{align}

In order to classify terms according to the small-$x$ expansion, and also to
compare with the known expressions~(\cref{cfrCatani,cfrIvanov,cfrBialas}) of impact
factors at leading order, it is convenient to expand the Mellin transforms of
the splitting functions as powers series in $\om$:
\begin{align}
  P_{qq}(\om) &= C_F \left(\frac54 -\frac{\pi^2}{3}\right)\om+\ord{\om^2}
   \label{Pqq}\\
  P_{gq}(\om) &= \frac{2 C_F}{\om}\left[1+\om A_{gq}(\om)\right] &
  A_{gq}(0) &= -\frac34  \label{Pgq}\\
  P_{qg}(\om) &= \frac23 T_R\left[1+\om A_{qg}(\om)\right] &
  A_{qg}(0) &=-\frac{13}{12} \label{Pqg}\\
  P_{gg}(\om) &= \frac{2 C_A}{\om}\left[1+\om A_{gg}(\om)\right] &
  A_{gg}(0) &= -\frac{11}{6} + \bbar \;, \quad
  \bbar = \frac{11}{12} - \frac{T_R N_f}{3 N_c} \label{Pgg}\\
  P_{q\gamma}(\om) &= \frac{N_c}{T_R}P_{qg}(\om) \;. \label{Pqgamma}
\end{align}
Note that $P_{qg}$ refers to the process where a gluon produces a single quark
emitting an antiquark, or viceversa. Therefore, a gluon splitting into a quark
or antiquark of a given flavour requires a factor of two. If the (anti-) quark at
some point splits into a gluon, the sum over flavours yields an additional
factor $N_f$. On the contrary, if the (anti-)quark couples to a photon, the sum
over flavours yields a factor $\sum_{q} e_q^2$. \Cref{Pqgamma} stems from
the fact that, if a gluon of colour $c$ splits into a quark-antiquark pair with
colours $a,b$, then the squared matrix element contains
$\sum_{ab}t^c_{ab}t^{d*}_{ab}=\tr(t^c t^d)=T_R \delta_{cd}$, while if a photon
splits into a quark-antiquark pair, the sum over colours is
$\sum_{ab}\delta_{ab}\delta_{ab}=\sum_a \delta_{aa}=N_c$.

By taking into account \cref{Pgq,Pqg,Pgg,Pqgamma} and noting that
$C_F N_c = (N_c^2-1)T_R$, we can rewrite \cref{TGT0} as
\begin{align}
  \dbtilde{\sigma}_0^{(TT)}(\om,\gamma;1)
  &= \left[ \al\as\Big(\sum_q e_q^2 \Big) 2 P_{qg}(\om) \sqrt{2(N_c^2-1)}
  \left(\frac1{\gamma^2}+\frac1{(1+\om-\gamma)^2}\right)
  \right]^2 \nonumber \\
   &\qquad\times  \frac1{\om}\,\Big(1+\om A_{gq}(\om)\Big)
   \; +\ord{\gamma^{-3}} + \ord{(1+\om-\gamma)^{-3}} \;,\label{Icoll}
\end{align}
where only the quartic poles in $\gamma$ are of our concern.

The term in square brackets is exactly the collinear limit of the
$\om$-dependent LO impact factor derived from \cref{cfrBialas}; in other
words, it represents the double poles of Bia\l as, Navelet and Peschanski (BNP) impact
factor~\cite{Bialas:2001ks} for a transverse photon with their
{\em full $\om$-dependent coefficient}:
\begin{align}\label{bnpT}
  \Phi_{\mathrm{BNP}}^{(T)}(\om,\gamma) &=
  \al\as\Big(\sum_q e_q^2\Big) T_R\sqrt{2(N_c^2-1)}\;
  \frac{\pi\Gamma(\gamma+\delta)\Gamma(\gamma)}{\Gamma(\omega) }\frac{1}{\left(\delta^2-1\right)\left(\delta^2-4\right)}
  \bigg\{\frac{\psi(\gamma + \delta) - \psi(\gamma)}{\delta} \times
  \nonumber\\
 &\qquad
  \frac{\omega^2\left[3(\om+1)^2+9\right]-2\om\left(\delta ^2 -1\right)+\left(\delta ^2 -1\right)\left(\delta ^2 -9\right)}{4\omega}-\frac{3(\omega+1)^2+3+\left(\delta^2-1\right)}{2}\bigg\} 
  \\
 &=  C_0\left[ \frac{1+\om A_{qg}}{\gamma^2}+ \frac{D(\om)}{\gamma}+\ord{\gamma^0}
  \right] + (\gamma\to 1+\om-\gamma) \label{bnpTcoll}\\
  C_0 &= \al\as\Big(\sum_q e_q^2\Big) \frac43 T_R\sqrt{2(N_c^2-1)}
  \label{C0} \;, \qquad
  D(\om) = \frac76 +\ord{\om}
   \;,\quad \delta\equiv \om+1-2\gamma \;.
\end{align}
The factor $1/\om$ --- stemming from $P_{gq}(\om)$ --- in the second
line of \cref{Icoll} yields the GGF~\eqref{rgiGGF} at lowest
order ($\as\to 0$), while the finite part $\propto A_{gq}$ provides a
NLL correction, to be reconsidered later.%
\footnote{In the $\om\to 0$ limit, \cref{Icoll} reduces to the product
of the LL GGF $1/\om$ with the LO impact factors
$\imf_0^{(T)}(\gamma)\imf_0^{(T)}(1-\gamma)$ of Catani et al. as in \cref{cfrCatani} --- restricting $\imf_0^{(T)}$ to the double poles in
$\gamma$.}
In conclusion,
\begin{equation}\label{collLO}
  \dbtilde{\sigma}_0^{(TT)}(\om,\gamma;1) = C_0^2\frac1{\om}\left[
  \frac{(1+\om A_{qg})^2 \,(1+\om A_{gq})}{\gamma^4}
  +\ord{\gamma^{-3}}\right] +(\gamma\to 1+\om-\gamma) \;.
\end{equation}

\subsection{LO RGI transverse impact factor\label{s:lotif}}

Our first task now is to determine (a possible form of) the LO RGI transverse
impact factor.  If we ignore for a moment the factor
$P_{gq}(\om)/P_{gq}(0)=(1+\om A_{gq})$,
\cref{Icoll} tells us that the LL $TT$ cross section {\em in the
  collinear limit} is given by the collinear limit of the LO
transverse impact factors of BNP~\cite{Bialas:2001ks},
times the LL GGF. One could then claim that the {\em full} LL RGI $TT$
cross section is given by the product of the complete BNP transverse
impact factors with the LL GGF, and conclude that
the LO RGI transverse impact factor is just the one provided in
ref.~\cite{Bialas:2001ks}, i.e., \cref{bnpT}. More properly,
such an impact factor is a perfect candidate, since it reproduces the LO BFKL cross 
section in the high-energy limit ($\om\to 0$) and also the LO DGLAP cross section
in the collinear limits $\gamma\to 0$ and $\gamma\to 1+\om$.

However, the collinear limit~\eqref{Icoll} of the cross section has the additional $\om$-dependent factor $P_{gq}(\om)/P_{gq}(0)$.
In order to take it into account, we
must modify either the BNP impact factors or the GGF. Since this factor stems
from the quark-gluon interaction, while the LL GGF is determined by
pure gluon dynamics, it is natural to associate such a factor to the
impact factors. The modification of impact factors is ambiguous, since
the collinear analysis just provides constraints for the leading twist
poles, i.e., for $\gamma\simeq 0$ and $\gamma\simeq 1+\om$, of their products.
Let's parametrize the leading-twist poles of $\Phi_0^{(T)}$ as follows:%
\footnote{We use the convention of parametrizing coefficients of the collinear and anti-collinear poles with the same letter, but with a bar over the coefficients of the anti-collinear poles.}
\begin{equation}\label{phi0col}
  \Phi_0^{(T)}(\om,\gamma;1)= C_0 
  \left[\frac{1+\om B(\om)}{\gamma^2}
    + \frac{D(\om)}{\gamma}
    +\frac{1+\om \bar{B}(\om)}{(1+\om-\gamma)^2}
    + \frac{\bar{D}(\om)}{1+\om-\gamma}\right]
    +r(\om,\gamma) \;,
\end{equation}
where $r(\om,\gamma)$ has no leading-twist poles. We then have, for $\gamma\simeq 0$,
\begin{align}\label{s0TT}
  \dbtilde{\sigma}_0^{(TT)}(\om,\gamma;1)
  &= \Phi_0^{(T)}(\om,\gamma;1) \;\frac{1}{\om}\;
  \Phi_{0}^{(T)}(\om,1+\om-\gamma;1) \nonumber \\
  &= C_0^2\frac1{\om}\left[
  \frac{(1+\om B)(1+\om \bar{B})}{\gamma^4}
  +\ord{\gamma^{-3}}\right] +(\gamma\to 1+\om-\gamma) \;.
\end{align}
Comparing the above expression with \cref{collLO} we get
\begin{subequations}\label{BpBt}
\begin{align}
  (1+\om B)(1+\om\bar{B}) &= (1+\om A_{qg})^2 (1+\om A_{gq}) \label{BpBte}\\
  \imp\quad B+\bar{B}&=2A_{qg}+A_{gq} + \ord{\om} \;. \label{BpBt0}
\end{align}
\end{subequations}
In the following, we often neglect the subleading terms $\ord{\om}$ in \cref{BpBt0}.

On the contrary, the coefficients $D(\om)$ and $\bar{D}(\om)$ of the simple
poles are out of reach of the present LO collinear analysis, but their value at
$\om=0$ can be determined from the explicit expression of \cref{phi0T}
in ref.~\cite{Catani:1994sq}: $D(0)=\bar{D}(0)=7/6$ [cfr.~\cref{C0}].  The
simplest and more natural choice for us is to adopt $D(\om)=\bar{D}(\om)$ as in
the impact factor $\Phi_{\mathrm{BNP}}^{(T)}$ of \cref{bnpT}.

According to the constraints previously derived, we present
some possible choices of the transverse LO RGI impact factor,
whose differences have to be considered a resummation-scheme ambiguity:
\footnote{Other schemes can be considered, see sec.~\ref{s:nttcscl}.}
\begin{subequations}\label{phiTchoice}
  \begin{align}
    \Phi_0^{(T)}(\om,\gamma;1) &=
    \Phi_{\mathrm{BNP}}^{(T)}(\om,\gamma)
    \left[1+\hom A_{gq}(\om)\right]
    &&(\text{scheme I}) \label{phiTa}\\
    \Phi_0^{(T)}(\om,\gamma;1) &=
    \Phi_{\mathrm{BNP}}^{(T)}(\om,\gamma)+C_0 \hom A_{gq}(\om)
    \left[\frac1{\gamma^2}+\frac1{(1+\om-\gamma)^2}\right]
    &&(\text{scheme II})  \label{phiTb}\\
    \Phi_0^{(T)}(\om,\gamma;1) &=
    \Phi_{\mathrm{BNP}}^{(T)}(\om,\gamma)
    +C_0\,\om A_{gq}(\om)\frac{1+\om A_{qg}}{(1+\om-\gamma)^2}
    &&(\text{scheme III})  \label{phiTc} \;.
  \end{align}
\end{subequations}
Scheme I is an overall renormalization of the impact factor. Scheme
II just modifies the coefficient of the (leading-twist) double poles.
Scheme III is motivated by the fact that the $P_{gq}$ vertex is attached to the impact factor to the right, thus providing a $1/\gamma$ pole only to $\Phi_0(\om,1-\gamma)$. Note that schemes I and II preserve the $\gamma\lra 1-\gamma$ symmetry of the impact factor, while scheme III does not. In particular $B=\bar{B}=A_{qg}+A_{gq}/2$ in schemes I and II, while $B=A_{qg}$, $\bar{B}=A_{qg}+A_{gq}+\om A_{qg}A_{gq}$ in scheme III (which fulfills exactly \cref{BpBte}).

\subsection{NLO $\boldsymbol{TT}$ cross section in the collinear limit\label{s:nttcscl}}

Our next task is to determine the transverse impact factors at NLO.
Specifically, we want
to determine a function $\Phi^{(T)}_1(\om,\gamma)$ such that
\begin{itemize}
\item
  the RGI cross section~\eqref{sigmaRGI} agrees with the NLL BFKL
  one~\eqref{sigmaBFKL};
\item the same RGI cross section agrees with the DGLAP cross section in the
  collinear limits $Q_1\gg Q_2$ and $Q_1\ll Q_2$ --- in this case at order
  $\as^3$.
\end{itemize}
The first condition has already been considered, and leads to the
constraint provided by \cref{F1expn} at $\om=0$.

\begin{figure}[ht]
  \centering
  \includegraphics[width=0.7\linewidth]{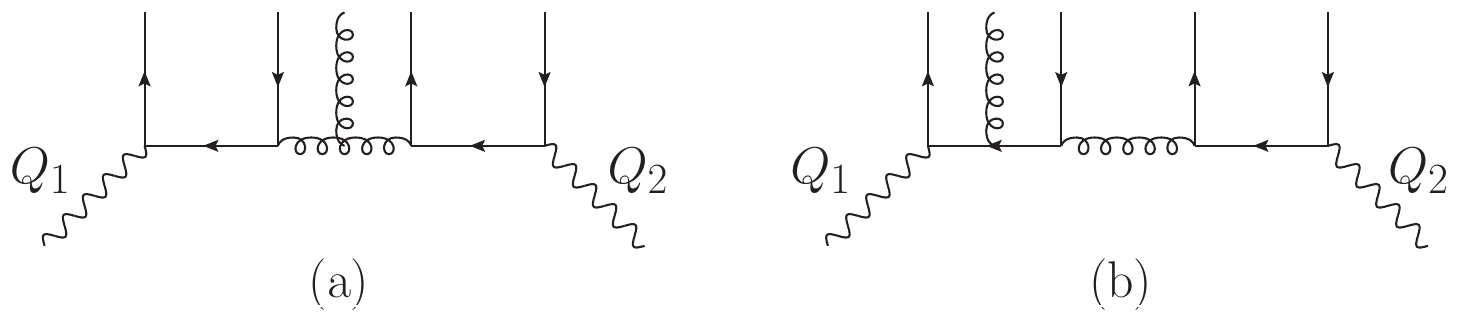}
  \caption{Ladder diagrams relevant in the collinear limit at next-to-leading order in the BFKL factorization formula. A third diagram, the left-right symmetric of (b) with the gluon emitted from the quark line on the right, is understood.  \label{f:colChainNL}}
\end{figure}

The second condition determines the structure of the collinear poles ($\gamma\simeq 0$ and $\gamma\simeq 1+\om$) of the impact factors.
We begin by generalizing \cref{TGT0} at $\ord{\as^3}$. This amounts to
consider ladder diagrams with five splittings between the photons. The vertices
at the photon legs are necessarily QED couplings as before. The other three
vertices are of QCD type, and lead to the diagrams of \cref{f:colChainNL} 
(a third diagram is understood, left-right symmetric with respect to (b) with the vertical gluon emitted from the quark on the right).
Such diagrams, together with the running-coupling term of \cref{rcCorr}, provide the integrand of the RGI factorization formula at $\ord{\as^3}$ in the collinear limit
\begin{equation}\label{colChainNLO}
  \as\,\dbtilde{\sigma}_1^{(TT)}(\om,\gamma;1)
  = \dbtilde{\sigma}_0^{(TT)}(\om,\gamma;1)
  \left[\frac{\as}{2\pi}\frac{P_{gg}}{\gamma}+2\frac{\as}{2\pi}\frac{P_{qq}}{\gamma}
  -\frac{\as b}{\gamma}+\ord{\gamma^0}\right] \;,
\end{equation}
where $\dbtilde{\sigma}_0^{(TT)}$ is the LO (collinear) integrand defined in \cref{colChainLO}.

The first term of \cref{colChainNLO} stems from the diagram of \cref{f:colChainNL}a which involves a $P_{gg}$ splitting function. According to the analysis of CCSS~\cite{Ciafaloni:2003rd}, this contribution can be entirely
associated to the GGF.
Actually, at fixed $\as$, the iteration of the $P_{gg}$ splitting function along the gluon ladder provides a geometric series that is easily summed, yielding
\begin{align}
  \om\,\ggf(\om,\gamma;1)^\coll &= \sum_{n=0}^\infty \left(\frac{\as}{2\pi}\frac{P_{gg}(\om)}{\gamma}\right)^n
  = \left[1-\frac{\asb}{\om}\frac{1+\om A_{gg}}{\gamma}\right]^{-1} \;. \label{omGcoll}
\end{align}
Since $\om\ggf = [1-\frac{\asb}{\om}X]^{-1}$
[cfr.~\cref{rgiGGF}], we find
$X_0^\coll(\om,\gamma;1)=[1+\om A_{gg}]/\gamma$.

The second term in \cref{colChainNLO} stems from the diagram of \cref{f:colChainNL}b and its symmetric counterpart --- with the gluon emitted from the quark line on the right --- which involve $P_{qq}$ splitting functions. It is naturally associated to the impact factors. Actually, since $P_{qq}$ vanishes at $\om=0$ (cfr.~\cref{Pqq}), these contributions are suppressed by two powers of $\om$ w.r.t.\ the diagram with $P_{gg}$, and thus are next-to-next-to-leading in the BFKL hierarchy. However, we keep them, in the spirit of being accurate in the leading DGLAP evolution.

The third term in \cref{colChainNLO} is the running coupling
($b$-dependent) contribution derived in \cref{rcCorr}, and can be incorporated into either the impact factors or the GGF, or both. In the following section, we face this situation more systematically, and propose some possible choices of transverse NLO RGI impact factor.

\subsection{NLO RGI transverse impact factor\label{s:nlotif}}

We now determine the NLO RGI impact factor from the NLO cross section derived in the previous section. Let's parametrize the collinear structure of RGI impact factors and kernel as follows (from now on in this section we suppress the superscript $(T)$ on the impact factors):
\begin{align}
    \Phi(\om,\gamma;1) &= \Phi_0(\om,\gamma;1) \left[1
    +\asb\left(\frac{M(\om)}{\gamma}+\frac{\bar{M}(\om)}{1+\om-\gamma}+r_1(\om,\gamma)\right)
    +\ord{\asb^2}\right] \label{phiTcol} \\
    X(\om,\gamma;1) &= \frac{1+\om U(\om)}{\gamma} + \ord{\gamma^0}
    +\asb\left(\frac{V(\om)}{\gamma^2}+\ord{\gamma^{-1}}\right)+\ord{\asb^2} \;,
    \label{Xcol}
\end{align}
where $r_1$ is regular at $\gamma=0,1+\om$ and we have taken into account that
additional powers of $\asb$ involve additional powers of $1/\gamma$ and
$1/(1+\om-\gamma)$.%
\footnote{We recall that $\Phi_0^{(T)}$ has collinear poles of second order. Therefore,
we expect an improved NL impact factor with cubic poles. This has to be contrasted with the
collinear behaviour of the BFKL impact factor $\imf_1^{(T)}$ featuring cubic and
even quartic poles at $\gamma=0,1$, as it is apparent from \cref{svilphi1}.
}
We find
\begin{align}
  \as\,\dbtilde{\sigma}_1(\om,\gamma;1)
  &= (\Phi_0\,\ggf_1\,\Phi_0 + \Phi_1\,\ggf_0\,\Phi_0 + \Phi_0\,\ggf_0\,\Phi_1)
  -\Phi_0\,\ggf_0\,\Phi_0 \nonumber \\
  &= \dbtilde{\sigma}_0(\om,\gamma;1)
  \;\asb\left[
  \frac{\frac1{\om}+M+\bar{M}+U}{\gamma}
  +\ord{\gamma^0} \right] \;. \label{I1TTrgi}
\end{align}
By comparing \cref{I1TTrgi} with \cref{colChainNLO} yields
\begin{equation}\label{MpMt}
  M+\bar{M}+U = A_{gg}+2 \bar{P}_{qq}-\bbar \;,
\end{equation}
having defined $\bbar$ in \cref{bbar}, $\bar{P}_{ab}\equiv P_{ab}/(2 C_A)$ and used $\bar{P}_{gg}=1/\om+A_{gg}$ [cfr.~\cref{Pgg}].

In order to check the compatibility of the collinear analysis with the known BFKL results, and also to further constrain the RGI impact factors, let us 
write down the collinear structure of the NLO BFKL transverse impact factor [\cref{phi0T,eq: Ivanov FT1}] and kernel~[\cref{chi0,eq:nllorg}].%
\footnote{With running-coupling scale $\mu_R^2=Q_1^2$, the double poles of $\chi_1(\gamma)$ are $A_1/\gamma^2$ and $(A_1-\bbar)/(1-\gamma)^2$. With symmetric scale $\mu_R^2=Q_1 Q_2$, the coefficients of both poles are equal to $A_1-\bbar/2$.}
At symmetric scales $s_0 = \mu_R^2 = Q_1 Q_2$, i.e., $p=0$:
\begin{align}
  \phi_0(\gamma) &= \phi_0(1-\gamma) = C_0\left(
  \frac1{\gamma^2}+\frac{D(0)}{\gamma}
  +\frac1{(1-\gamma)^2}+\frac{\bar{D}(0)}{1-\gamma} + \cdots\right) \label{svilphi0}\\
  \phi_1(\gamma) &= \phi_0(\gamma)\left(\frac{-1}{\gamma^2}+\frac{\eta}{\gamma}
  +\frac{-3/2}{(1-\gamma)^2}+\frac{\bar{\eta}}{1-\gamma}
  +\cdots\right) \label{svilphi1} \\
  \chi(\gamma) &= \frac1{\gamma} + \asb\left(\frac{-1/2}{\gamma^3}+\frac{A_1(0)-\bbar/2}{\gamma^2}+\frac{H_1}{\gamma}+\cdots\right) \label{svilchi} \\
  &\eta = -\frac{11}{6}\;, \quad \bar{\eta} = -\frac{7}{4}
  \;, \quad H_1 = -\frac{T_R N_f}{N_c}\left(\frac59+\frac{13}{18 N_c^2}\right) \;, \label{eta}
\end{align}
where $C_0$, $D(0)$ and $\bar{D}(0)$ were already determined in \cref{C0} and $A_1$ was defined in \cref{A1}.
Then, from \cref{chi1expn} we find
\begin{equation}\label{UpV}
  A_1-b/2 = U+V \;.
\end{equation}
By noting that $\imf_0(\gamma)=\imf_0(1-\gamma)=\Phi_0(0,\gamma;p)$ for any $p$, \cref{F1expn} can be simplified into
\begin{align}
  \imf_1(\gamma)+\imf_1(1-\gamma) &=
  \Phi_1(0,\gamma)+ \Phi_1(0,1-\gamma) \nonumber \\
  &\quad+\chi_0(\gamma)[
  \de_\om \Phi_0(0,\gamma)+\de_\om \Phi_0(0,1-\gamma)]+\imf_0(\gamma)\de_\om X_0(0,\gamma)\;, \label{sumIfT}
\end{align}
where $\Phi$'s and $X$ must be considered here at $p=0$, i.e., by replacing $\gamma\to\gamma+\om/2$ in \cref{phiTcol,Xcol}.
From \cref{svilphi1} we can expand the l.h.s.\ of \cref{sumIfT} around the collinear pole $\gamma=0$:
\begin{equation}
  \imf_1(\gamma)+\imf_1(1-\gamma) = \phi_0(\gamma)\;
  \asb\left[\frac{-5/2}{\gamma^2}+\frac{\eta+\bar{\eta}}{\gamma}
    +\ord{\gamma^0} \right]  \;.
  \label{medimf1}
\end{equation}
By expanding the r.h.s.\ of \cref{sumIfT} using \cref{phiTcol,Xcol}
--- with the replacement $\gamma\to\gamma+\om/2$ ---
, the coefficient $-5/2$ of the quadratic pole within square brackets in \cref{medimf1} is correctly reproduced, while the coefficients of the simple poles are equal if
\begin{equation}\label{etaCond}
  \eta+\bar{\eta} = B+\bar{B}+\half D+\half \bar{D}+M+\bar{M}+U\big|_{\om=0} \;.
\end{equation}
This is indeed the case. In fact, by exploiting \cref{BpBt} and \cref{MpMt}, we find
\begin{subequations}
\begin{align}
  B+\bar{B}+\half D+\half \bar{D}+M+\bar{M}+U\big|_{\om=0}
  &= 2\bar{P}_{qq}+2A_{qg}+A_{gq}+A_{gg}+\half D+\half \bar{D}-\bbar \big|_{\om=0} \label{BpMpU} \\
 &= -\frac{43}{12} =\eta+\bar{\eta} \;,
\end{align}
\end{subequations}
thus proving the consistency of next-to-leading BFKL and leading-order DGLAP.

Of course, the constraints \eqref{UpV} and \eqref{BpMpU} derived from \cref{chi1expn,F1expn} respectively, can be fulfilled in many ways.
In \cref{t:Phi1Tschemes} we present some choices that we prefer on physical grounds.

\begin{table}[htp]
    \centering
\begin{tabular}{|l|r|r|r|r|}
\hline
  scheme name & $U$            & $V$  & $B+\bar{B}$ & $M+\bar{M}$ \\
 \hline
  collA       & $A_{gg}-\bbar$ & $\Delta A + \bbar/2$ &  $2A_{qg}+A_{gq}$ & $2 \bar{P}_{qq}$ \\
  collB       & $A_{gg}$       & $\Delta A - \bbar/2$ & $2A_{qg}+A_{gq}$  & $2\bar{P}_{qq}-\bbar$ \\
  zVnB        & $A_1-\bbar/2$  & $0$                  & $2A_{qg}+A_{gq}$  & $2\bar{P}_{qq}-\Delta A-\bbar/2$ \\
  zVnM        & $A_1-\bbar/2$  & $0$                  & $2A_{qg}+A_{gq}-\Delta A - \bbar/2$ & $2\bar{P}_{qq}$ \\
  zVzM        & $A_1-\bbar/2$  & $0$                  & $2\bar{P}_{qq}+2A_{qg}+A_{gq}-\Delta A-\bbar/2$  & $0$ \\
  \hline
\end{tabular}
   \caption{Favourite scheme choices for defining the NLO RGI transverse impact factor.}
\label{t:Phi1Tschemes}
\end{table}

Schemes ``collA" and ``collB"  are motivated by the collinear analysis that suggests the value of $B+\bar{B}$ from \cref{BpBt} and the values of $M+\bar{M}$ and $U$ from
\cref{MpMt,omGcoll}. In the former we assign the running-coupling term $-\bbar$ to the kernel, in the latter to the impact factors. For convenience, we have introduced
\begin{equation}\label{deltaA}
  \Delta A = A_1(\om) - A_{gg}(\om) = \frac{C_F}{C_A} 2N_f\bar{P}_{qg}(0) = \left(1-\frac1{N_c^2}\right)\frac{T_R N_f}{3 N_c} \;.
\end{equation}

In the other three schemes ``zV\dots" we set to zero the coefficient $V$ of the double pole of $X_1$, following the spirit of the RG improvement to transfer the most singular $\gamma$-poles of NL objects into regular $\om$-corrections of leading-order terms. In this way, we assign all the dependence of the kernel on the gluon anomalous dimension and running coupling $A_1-\bbar/2$ to the $\ord{\om}$-term of the leading eigenvalue $X_0$.
Scheme ``zVnB" adopts the natural (i.e., collinearly motivated) choice for the $B$'s coefficients; scheme ``zVnM" adopts the natural choice for the $M$'s coefficients; scheme ``zVzM" sets to zero the coefficients $M$ of the cubic poles of the NLO impact factors, thus assigning all the residual dependence on the anomalous dimensions to the $\ord{\om}$-term of the leading impact factor $\Phi_0$.

Actually, each of the schemes in \cref{t:Phi1Tschemes} can be implemented in many ways, depending on how $B$, $\bar{B}$, $M$ and $\bar{M}$ are individually defined, and also because the regular part of impact factors is fully constrained only at $\om=0$. Concerning the leading impact factor $\Phi_0^{(T)}$, we propose the three sub-schemes of \cref{phiTchoice}, where $B=\bar{B}$ in the sub-schemes I and II, while $\bar{B}=B+A_{gq}$ in sub-scheme III.

As for the leading eigenvalue function, we adopt the recipe proposed in ref.~\cite{Ciafaloni:2003rd}: 
\begin{equation}\label{X0}
  X_0(\om,\gamma;0) =  2\psi(1) - \psi(\gamma + \frac{\omega}{2}) - \psi(1-\gamma+ \frac{\omega}{2}) + \omega U(\om) \left( \frac{1}{\gamma + \omega/2} + \frac{1}{1-\gamma + \omega/2} \right) \;,
\end{equation}
where $U(\om)$, according to \cref{t:Phi1Tschemes}, depends on the scheme choice. Then, according to \cref{chi1expn}, the next-to-leading improved eigenvalue at $\om=0$ reads
\begin{equation}
    X_1(0,\gamma) = \chi_1(\gamma) + \frac{1}{2}\chi_0(\gamma)\frac{\pi^2}{\sin^2 \pi \gamma}- U(0)\left(\frac1{\gamma}+\frac1{1-\gamma}\right)\chi_0(\gamma) \;.
    \label{eq:NLOsubtractions0}
\end{equation}
The above expression is free of cubic poles, but still contains simple poles and possibly double poles, depending on the scheme choice:
\begin{equation}
  X_1(0,\gamma) = \left(A_1-\frac{\bbar}{2}-U(0)\right)\frac1{\gamma^2}
  +\left(H_1+\frac{\pi^2}{6}-U(0)\right)\frac{1}{\gamma} + \cdots \;.
\end{equation}
According to the RGI method, we require the RGI eigenvalue function $X_1(\om,\gamma)$ to have poles at the expected $\om$-shifted positions. The final expression that we adopt is
\begin{align}
    X_1(\om,\gamma) = X_1(0,\gamma) 
    &+\frac{A_1(\om)-\frac{\bbar}{2}-U(\om)}{(\gamma+\hom)^2}-\frac{A_1(0)-\frac{\bbar}{2}-U(0)}{\gamma^2}
    +(\gamma\lra 1-\gamma)
    + \nonumber \\
    &+\left(H_1+\frac{\pi^2}{6}
    -U(0)\right)\left[X_0(\omega,\gamma)-\chi_0(\gamma)\right] \;.
    \label{eq:NLO subtractions}
\end{align}

We can now exploit \cref{sumIfT} to constrain the NLO improved transverse impact factor at $\om=0$ and arbitrary $\gamma$. If we
further require such impact factor to be symmetric in $\gamma\to 1-\gamma$, we obtain%
\footnote{We note that, while $\phi_0(\gamma)$ is symmetric in $\gamma\to 1-\gamma$,
the NLO impact factor $\phi_1(\gamma)$ is not. Actually, since the latter has been
derived~\cite{Ivanov:2014hpa} from a cross section~\cite{Chirilli:2013kca} which depends
on the product $\imf^{(T)}(\gamma)\imf^{(T)}(1-\gamma)$, it is not clear to us how
$\imf_1$ has been unambiguously derived, without imposing further requirements.
}
\begin{align}
  \Phi_1(0,\gamma) &=
  \frac12\left[\Phi_1(0,\gamma)+\Phi_1(0,1-\gamma)\right] \nonumber \\
  &= \frac12\left[\imf_1(\gamma)+\imf_1(1-\gamma)
   -\imf_0(\gamma)\de_\om X_0(0,\gamma)
  -\chi_0(\gamma) \big( \de_\om \Phi_0(0,\gamma)
  +\de_\om \Phi_0(0,1-\gamma) \big) \right]\;.
  \label{PhiT1om0}
\end{align}
Its Laurent expansion around $\gamma=0$ reads
\begin{equation}\label{laurentT}
  \Phi_1(0,\gamma) = C_0 \left[\frac{M(0)}{\gamma^3}
  +\frac{M_2}{\gamma^2}+\frac{M_1}{\gamma} +\ord{\gamma^0}\right]\;,
\end{equation}
where $M_2$ and $M_1$ depend on the scheme choice that defines the $\om$-dependence of $\Phi_0(\om,\gamma)$ and $X_0(\om,\gamma)$ in \cref{phiTchoice,X0}.

We extend $\Phi_1$ at $\om\neq 0$ by requiring the collinear poles to be located at $\gamma=-\om/2$ and $\gamma=1+\om/2$ and with $\om$-dependent leading coefficients $M(\om)$ and $\bar{M}(\om)$ as in \cref{phiTcol}. This can be obtained in various ways, and we adopt the following choice:
\begin{align}
  \Phi_1(\om,\gamma;0) &= \Phi_1(0,\gamma) \nonumber \\
  &+C_0\left\{\left[\frac{M(\om)}{(\gamma+\hom)^3}
  +\frac{M_2}{(\gamma+\hom)^2}+\frac{M_1}{\gamma+\hom}\right]
  -\left[\frac{M(0)}{\gamma^3}
  +\frac{M_2}{\gamma^2}+\frac{M_1}{\gamma}\right]
  +\begin{pmatrix} \gamma\lra 1-\gamma \\ M\to\bar{M} \end{pmatrix}
  \right\} \;. \label{phiT1om}
\end{align}
Having required $\Phi_1$ to be symmetric causes $M(\om)=\bar{M}(\om)$ equal to
half the expression in the last column of \cref{t:Phi1Tschemes}.

\section{RGI impact factor for longitudinal photons\label{s:rgilif}}

\subsection{Cross section and impact factor at leading order\label{s:ifLO}}

In order to determine the longitudinal RGI impact factor at leading order, we first
consider the cross section $\sigma^{(LT)}(Q_1,Q_2)$
where the photon $Q_1$ (on the left) has longitudinal polarization, while the other one
$Q_2$ (on the right) is transverse. We are interested in the collinear limit $Q_1^2 \gg Q_2^2$,
therefore we need the vertices that describe how the longitudinal
photon $Q_1$ couples to quarks and gluons $k$ in the collinear limit
$Q_1^2\gg k^2$. They can be derived from the longitudinal coefficient functions,
as explained in \cref{s:locs}.

\begin{figure}[ht]
  \centering
  \includegraphics[width=0.3\linewidth]{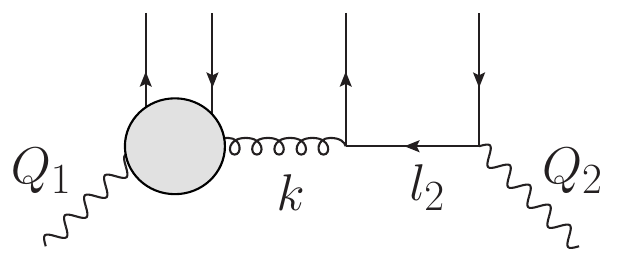}
  \caption{Diagramatics of collinear limit at leading order for the longitudinal impact factor. The blob represents the gluonic contribution to the longitudinal coefficient function at lowest order in $\as$.}\label{f:longColChain}
\end{figure}

The lowest order ladder diagram of this kind involving a high-energy gluon exchange is
depicted in \cref{f:longColChain}.
The shaded circle at
the left represents the gluon contribution to the longitudinal coefficient function $C_L^g$, while the two vertices on the right side represent two splitting functions, as in the $TT$-case.

We can then repeat the collinear analysis of sec.~\ref{s:cscl} by replacing in \cref{colChainLO} the ``transverse" factor eq.  \eqref{ITg1} with the ``longitudinal" factor eq. \eqref{ILg1} (see Appendix), thus obtaining
the leading $\gamma$-pole structure of $\dbtilde{\sigma}^{(LT)}$:
\begin{align}
  \dbtilde{\sigma}_0^{(LT)}(\om,\gamma;1)_{\coll} &=
  \frac{\alpha\as \big(\sum_q e_q^2\big) 8 T_R\sqrt{2(N_c^2-1)}}{
    \gamma\,(2+\om)(3+\om)}
  \;\frac{1+\om A_{gq}(\om)}{\om}\;
  \frac{\alpha\as \big(\sum_q e_q^2\big)
    2 P_{qg}(\om)\sqrt{2(N_c^2-1)}}{\gamma^2}
  +\ordd{\gamma^{-2}} \nonumber \\
  &=\Phi_{\mathrm{BNP}}^{(L)}(\om,\gamma)
    \;\frac{1+\om A_{gq}(\om)}{\om}\;
    \Phi_{\mathrm{BNP}}^{(T)}(\om,1+\om-\gamma)
    +\ordd{\gamma^{-2}} \;,\label{IcollLT}
\end{align}
namely the product of the corresponding BNP impact factors with exact
kinematics~\cite{Bialas:2001ks}, the LO GGF $1/\om$ and the same $\ord{\om}$ correction $\propto A_{gq}(\om)$.
The second line of \cref{IcollLT} follows from the collinear structure of the BNP impact factors, reported in \cref{bnpTcoll} for the transverse polarization and in the following equation for the longitudinal polarization:
\begin{align}
  \Phi_{\mathrm{BNP}}^{(L)}(\om,\gamma) &=
  \al\as\Big(\sum_q e_q^2\Big) T_R\sqrt{2(N_c^2-1)}\;4
 \frac{\pi\Gamma(\gamma+\delta+1)\Gamma(\gamma+1)}{\Gamma(\omega)}\frac{1}{\left(\delta^2-1\right)\left(\delta^2-4\right)}
 \times \nonumber \\
 &\qquad \left[\frac{\psi(\gamma + \delta) - \psi(\gamma)}{\delta}\cdot \frac{3\omega^2-\left(\delta^2-1\right)}{2\omega}-3\right] \label{bnpL} \\
  &= C_0 \left[\frac{1+\om\Lambda(\om)}{\gamma} + D_L(\om)
    +\ord{\gamma}\;+\;(\gamma\lra 1+\om-\gamma)\right] \label{bnpLcol} \\
  1+\om\Lambda(\om)&=\frac{6}{(2+\om)(3+\om)}\;, \quad
  \Lambda(0)=-\frac{5}{6} \;, \quad D_L(0) = -\frac13
  \;,\quad \delta\equiv 1+\om-2\gamma \;,
  \label{Lam0}
\end{align}
where $C_0$ is the same normalization coefficient of the transverse impact factor,
as given in \cref{C0}.
Therefore, \cref{IcollLT} can be rewritten as
\begin{equation}\label{sig0LT}
  \dbtilde{\sigma}_0^{(LT)}(\om,\gamma;1) = C_0^2\frac1{\om}\left[
    \frac{(1+\om\Lambda)(1+\om A_{gq})(1+\om A_{qg})}{\gamma^3}
    +\ord{\gamma^{-2}} \right] \;.
\end{equation}

Taking inspiration from \cref{phi0col,bnpLcol}, we parametrize the collinear structure of the longitudinal LO RGI impact factor as
\begin{equation}\label{phi0Lcol}
  \Phi_0^{(L)}(\om,\gamma;1)= C_0 
  \left[\frac{1+\om B_L(\om)}{\gamma}
    + D_L(\om)
    +\frac{1+\om \bar{B}_L(\om)}{1+\om-\gamma}
    + \bar{D}_L(\om) \right]
    +r_L(\om,\gamma) \;,
\end{equation}
where $r_L(\om,\gamma)$ vanishes at $\gamma=0$ and $\gamma=1+\om$. By combining the above expression with the analogue one in \cref{phi0col}, we find
\begin{align}
  \dbtilde{\sigma}^{(LT)}_{0}(\om,\gamma;1)
  &= \Phi_0^{(L)}(\om,\gamma;1) \;\frac{1}{\om}\;
  \Phi_{0}^{(T)}(\om,1+\om-\gamma;1) \nonumber \\
  &= C_0^2\frac1{\om}\left[
  \frac{(1+\om B_L)(1+\om\bar{B})}{\gamma^3}
  +\ord{\gamma^{-2}}\right] +(\gamma\to 1+\om-\gamma) \;. \label{sigma0LT}
\end{align}
If we compare \cref{sigma0LT} with \cref{sig0LT}, we obtain a relation among $B_L$, $\bar{B}$ and the known quantities $\Lambda, A_{qg}, A_{gq}$. However, remembering that $B$ and $\bar{B}$ are constrained by \cref{BpBt}, we can actually relate $B_L$ and $B$:
\begin{align}
  \frac{1+\om B_L}{1+\om B} &= \frac{1+\om\Lambda}{1+\om A_{qg}} \label{BLoB}\\
  \imp\quad B_L &= \Lambda + B - A_{qg} + \ord{\om} \;. \label{BLrel}
\end{align}

The coefficient $\bar{B}_L$ of the simple anti-collinear pole can be determined in an analogous way by considering the cross section for two longitudinal photons, i.e., by comparing the two expansions for
\begin{align}
  \dbtilde{\sigma}^{(LL)}_{0}(\om,\gamma;1)
  &= \Phi_0^{(L)}(\om,\gamma;1) \;\frac{1}{\om}\;
  \Phi_{0}^{(L)}(\om,1+\om-\gamma;1) \nonumber \\
  &=\Phi_{\mathrm{BNP}}^{(L)}(\om,\gamma)
    \;\frac{1+\om A_{gq}(\om)}{\om}\;
    \Phi_{\mathrm{BNP}}^{(L)}(\om,1+\om-\gamma)
    +\ordd{\gamma^{-1}} \;, \label{sigmaLL}
\end{align}
yielding
\begin{subequations}\label{BLbarBL}
\begin{align}
  (1+\om B_L)(1+\om\bar{B}_L) &= (1+\om\Lambda)^2 (1+\om A_{gq}) \label{BLbarBLe} \\
  \imp\quad B_L + \bar{B}_L &= 2\Lambda + A_{gq}+\ord{\om} \;. \label{BLbarBL0}
\end{align}
\end{subequations}
Note that the role played by $A_{qg}$ for $B$ and $\bar{B}$ in \cref{BpBt} is now played by $\Lambda$ for $B_L$ and $\bar{B}_L$ in \cref{BLbarBL}.

Just like in the transverse case, we
can define the LO RGI longitudinal impact factor by sharing
the $A_{gq}$ correction term between the leading collinear and anticollinear poles:
\begin{subequations}\label{phiLchoice}
  \begin{align}
    \Phi_0^{(L)}(\om,\gamma;1) &=
    \Phi_{\mathrm{BNP}}^{(L)}(\om,\gamma)
    \left[1+\hom A_{gq}(\om)\right]
    &&(\text{scheme I}) \label{phiLa}\\
    \Phi_0^{(L)}(\om,\gamma;1) &=
    \Phi_{\mathrm{BNP}}^{(L)}(\om,\gamma)+C_0 \hom A_{gq}(\om)
    \left[\frac1{\gamma}+\frac1{1+\om-\gamma}\right]
    &&(\text{scheme II})  \label{phiLb}\\
    \Phi_0^{(L)}(\om,\gamma;1) &=
    \Phi_{\mathrm{BNP}}^{(L)}(\om,\gamma)
    +C_0\,\om A_{gq}(\om)\frac{1+\om A_{qg}}{1+\om-\gamma}
    &&(\text{scheme III})  \label{phiLc} \;.
  \end{align}
\end{subequations}
Schemes I and II implement the choice $B_L=\bar{B}_L=\Lambda+A_{gq}/2$, giving rise to symmetric impact factors, while scheme III has $B_L=\Lambda$ and
$\bar{B}_L=\Lambda+A_{gq}+\om\Lambda A_{gq}$, giving rise to an asymmetric impact factor, but fulfilling exactly \cref{BLbarBLe}.

\subsection{Cross section and impact factor at next-to-leading order\label{s:ifNLO}}

\begin{figure}[htp]
  \centering
  \includegraphics[width=0.9\linewidth]{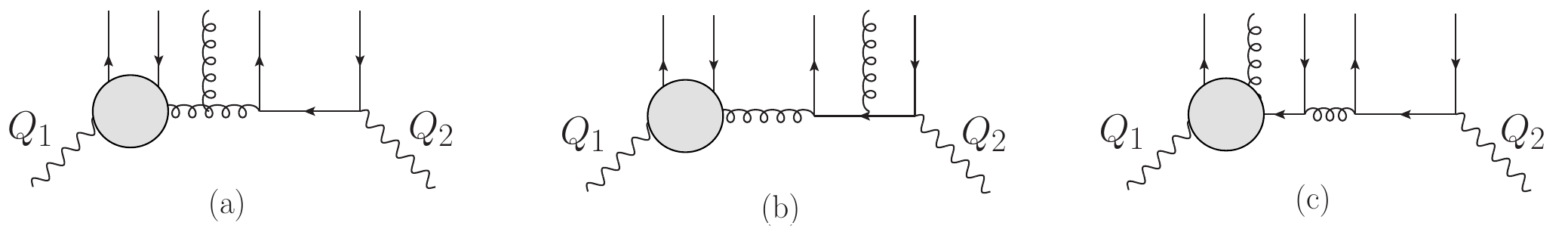}
  \caption{Diagramatics of collinear limit at next-to-leading order for the
    longitudinal impact factor.
    (a) Photon-gluon coefficient function and gluon emission from gluon line;
    (b) Photon-gluon coefficient function and gluon emission from quark line.
    (c) Photon-quark coefficient function and quark emission from parent gluon;
    \label{f:longColChainNLO}}
\end{figure}

The collinear analysis at NLO for the longitudinal-transverse photon cross section involves the three diagrams depicted in \cref{f:longColChainNLO} and can be presented in the following form:
\begin{equation}\label{sig1LT}
  \as\,\dbtilde{\sigma}_1^{(LT)}(\om,\gamma;1)
  = \dbtilde{\sigma}_0^{(LT)}(\om,\gamma;1)
  \left[\frac{\as}{2\pi}\frac{P_{gg}}{\gamma}
  +\frac{\as}{2\pi}\frac{P_{qq}}{\gamma}
  +\frac{C_F}{T_R}\frac{3+\om}{2}\cdot \frac{\as}{2\pi}\frac{P_{qg}}{\gamma}
  -\frac{\as b}{\gamma}+\ord{\gamma^0}\right] \;,
\end{equation}
where $\dbtilde{\sigma}_0^{(LT)}$ is the LO integrand defined in \cref{sig0LT}.

The first term in the r.h.s. of \cref{sig1LT} stems from the diagram of \cref{f:longColChainNLO}(a) involving a $P_{gg}$ splitting function, and can be entirely associated to the GGF.

The second term 
stems from the diagram of \cref{f:longColChainNLO}(b), with a gluon emitted from the quark
line on the right, and it is naturally associated to the
impact factor of the transverse photon $Q_2$.

The third term 
stems from the diagram of \cref{f:longColChainNLO}(c), which is genuinely different from other diagrams, because it involves a coefficient function where the longitudinal photon $Q_1$ couples to a quark. As explained in \cref{s:loph}, the photon-quark coefficient function differs from the photon-gluon one by the multiplicative factor $C_F(3+\om)/(2 T_R)$ [cfr.~\cref{LqLgFact}]; just to the right of the blob, we find the vertex with the $P_{qg}$ splitting function.
This contribution is naturally associated to the impact factor of the longitudinal photon $Q_1$.

The fourth and last term in \cref{sig1LT} is the running coupling
($b$-dependent) contribution derived in \cref{rcCorr}, and can be
incorporated into either the impact factors or the GGF, or both.

In order to determine the NLO RGI longitudinal impact factor from the NLO cross section,
we parametrize the collinear structure of the longitudinal impact factor exactly as in \cref{phiTcol}, by appending the subscript $L$ to the various (unbarred) coefficients,
e.g., $M\to M_L$. A straightforward calculation yields
\begin{equation}
  \as\,\dbtilde{\sigma}_1^{(LT)}(\om,\gamma;1)
  = \dbtilde{\sigma}_0^{(LT)}(\om,\gamma;1)
  \;\asb\left[
  \frac{\frac1{\om}+M_L+\bar{M}+U}{\gamma}
  +\ord{\gamma^0} \right] \;. \label{I1LTrgi}
\end{equation}
which is nothing but the result of \cref{I1TTrgi} with $T\to L$ in the first impact factor. We then derive [cfr.~\cref{MpMt} and the subsequent definitions]
\begin{align}
  M_L+\bar{M}+U &= \plq + \bar{P}_{qq} + A_{gg}-\bbar \;, \qquad
  \plq(\om) \equiv \frac{C_F}{T_R} \bar{P}_{qg}\frac{3+\om}{2} \label{MLpMt}\\
  \imp\quad M_L - M &= \plq - \bar{P}_{qq}
  = \frac{C_F}{2 C_A}+\ord{\om} \;. \label{MLmM}
\end{align}

In order to check the compatibility of the collinear analysis with the known BFKL results,
let us write down the collinear structure of the NLO BFKL longitudinal impact factors [\cref{phi0L,eq: Ivanov FL1}]:
\begin{align}
  \phi_0^{(L)}(\gamma) &= \phi_0^{(L)}(1-\gamma) = C_0\left(
  \frac1{\gamma} + D_L(0) + \ord{\gamma}\right) \label{svilphi0L}\\
  \phi_1^{(L)}(\gamma) &= \phi_0^{(L)}(\gamma)\left(\frac{-1/2}{\gamma^2}+\frac{\eta_L}{\gamma}
  +\frac{-1}{(1-\gamma)^2}+\frac{\bar{\eta}_L}{1-\gamma}
  +\cdots\right) \;, \quad
   \label{svilphi1L}
   \eta_L = -\frac{7}{3}\;, \quad \bar{\eta}_L = -\frac{9}{4} \;.
\end{align}
At this point we expand \cref{F1expn} around $\gamma=0$ by using \cref{svilphi0,svilphi1,svilchi,eta,svilphi0L,svilphi1L} for the LHS and \cref{phiTcol,Xcol} for the RHS. As a result, the (spurious) leading $\gamma$ poles match, while the subleading ones are equal if
\begin{equation}\label{etaLTCond}
  \eta_L+\bar{\eta} = B_L+\bar{B}+\half (D_L+ \bar{D}) +M_L+\bar{M}+U\big|_{\om=0} \;.
\end{equation}
Subtracting \ref{BpMpU} from the above equation, we obtain a relation among collinear coefficients:
\begin{equation}\label{etaDiff}
  \eta_L-\eta = B_L-B + \half(D_L-D) + M_L-M \;.
\end{equation}
While the LHS evaluates to $-1/2$,
the RHS is equal to $-1/2 + C_F/(2 C_A)$. Therefore, we find agreement with the result of ref.~\cite{Ivanov:2014hpa}, were it not for the presence of a
term proportional to $C_F$ Casimir in the collinear pole. It looks like their impact factor misses the contribution from the diagram of \cref{f:longColChainNLO}c.

Finally, by considering the cross section for two longitudinally polarized cross section, we obtain the result of \cref{etaLTCond} with the barred (transverse) coefficients replaced by their corresponding longitudinal counterparts:
\begin{equation}\label{etaLLCond}
  \eta_L+\bar{\eta}_L = B_L+\bar{B}_L+\half (D_L+\bar{D}_L)+M_L+\bar{M}_L+U\big|_{\om=0} \;,
\end{equation}
which is satisfied only if $M_L+\bar{M}_L=\ord{\om}$. If we take $\bar{M}_L=M_L$, as it is natural to assume in the transverse case, then we disagree with the result of ref.~\cite{Ivanov:2014hpa} by a $C_F$ term in the $1/\gamma$ pole of the ratio $\imf_1^{(L)}/\imf_0^{(L)}$.

We collect our results for the longitudinal RGI impact factor in \cref{t:Phi1Lschemes}.

\begin{table}[htp]
    \centering
\begin{tabular}{|l|r|r|r|r|}
\hline
  scheme name & $U$            & $V$  & $B_L+\bar{B}_L$ & $M_L+\bar{M}_L$ \\
 \hline
  collA       & $A_{gg}-\bbar$ & $\Delta A + \bbar/2$ &  $2\Lambda+A_{gq}$ & $2 \plq$ \\
  collB       & $A_{gg}$       & $\Delta A - \bbar/2$ & $2\Lambda+A_{gq}$  & $2\plq-\bbar$ \\
  zVnB        & $A_1-\bbar/2$  & $0$                  & $2\Lambda+A_{gq}$  & $2\plq-\Delta A-\bbar/2$ \\
  zVnM        & $A_1-\bbar/2$  & $0$                  & $2\Lambda+A_{gq}-\Delta A - \bbar/2$ & $2\plq$ \\
  zVzM        & $A_1-\bbar/2$  & $0$                  & $2\plq+2\Lambda+A_{gq}-\Delta A-\bbar/2$  & $0$ \\
  \hline
\end{tabular}
   \caption{Favourite scheme choices for defining the NLO RGI longitudinal impact factor. The values of $U$ and $V$ are the same as in \cref{t:Phi1Tschemes}.}
\label{t:Phi1Lschemes}
\end{table}

Once a scheme has been chosen, the LO impact factor $\Phi_0^{(L)}$ can be
specified according to one of the sub-schemes in \cref{phiLchoice}, with
$B_L=\bar{B}_L$ in sub-schemes I and II, while $\bar{B}_L=B_L+A_{gq}$ in
sub-scheme III.

The NLO impact factor $\Phi_1^{(L)}$ is constructed to be symmetric, as in the transverse case: at $\om=0$ \cref{PhiT1om0} holds unaltered,
provided we add to $\imf_1$ the contribution 
\begin{equation}
    \Delta \imf_1^{(L)}(\gamma)=\plq(0)\left(\frac{1}{\gamma}+\frac{1}{1-\gamma}\right) \imf_0^{(L)}(\gamma)\;.
\end{equation}
The Laurent expansion around $\gamma=0$ shows a double pole
\begin{equation}\label{laurentL}
  \Phi_1^{(L)}(0,\gamma) = C_0 \left[\frac{M_L(0)}{\gamma^2}
  +\frac{M_{L,1}}{\gamma} +\ord{\gamma^0}\right]\;,
\end{equation}
where $M_{L,1}$ depends on the scheme choice.
We extend $\Phi_1$ at $\om\neq 0$ by requiring the collinear poles to be located at $\gamma=-\om/2$ and $\gamma=1+\om/2$ and with $\om$-dependent leading coefficients $M_L(\om)$ and $\bar{M}_L(\om)$ as we did for the transverse impact factor:
\begin{align}
  \Phi_1^{(L)}(\om,\gamma;0) &= \Phi_1^{(L)}(0,\gamma) \nonumber \\
  &+C_0\left\{\left[\frac{M_L(\om)}{(\gamma+\hom)^2}
  +\frac{M_{L,1}}{\gamma+\hom}\right]
  -\left[\frac{M_L(0)}{\gamma^2}
  +\frac{M_{L,1}}{\gamma}\right]
  +\begin{pmatrix} \gamma\lra 1-\gamma \\ M\to\bar{M} \end{pmatrix}
  \right\} \;. \label{phiL1om}
\end{align}
Having required $\Phi_1^{(L)}$ to be symmetric causes $M_L(\om)=\bar{M}_L(\om)$ equal to
half the expression in the last column of \cref{t:Phi1Lschemes}.

\section{Numerical analysis}
\label{s:numerical}

In this section, we apply the factorization formula with renormalization-group improved impact factors and Green's function to compute the $\gamma^* \gamma^*$ cross section in phenomenologically relevant situations.
The presented results contain the sum over all combinations of photon polarizations:
$\sigma = \sigma^{(TT)}+ \sigma^{(LT)}+ \sigma^{(TL)}+ \sigma^{(LL)}$.
For the NLL RGI  calculation, the $\sigma^{(TT)}$ is about 56\% of the total cross section on average at  $Q^2=17 \,\rm GeV^2$, while both $\sigma^{(TL)}$ and $\sigma^{(LT)}$ about 19\%, $\sigma^{(LL)}$ about 6\%. 
These percentages vary by about $2\%$ for $\sigma^{(TT)}$, and about $1\%$ for other polarization combinations upon changes of the scheme and varying rapidity between 2 and 7 units. 

The numerical calculation is based on the following formulae:
the cross section is calculated using \cref{sigmaRGI2};
the leading eigenvalue $X_0$ is given in \cref{X0};
the NL one $X_1$ in \cref{eq:NLOsubtractions0,eq:NLO subtractions} with $\chi_1$ in \cref{eq:nllorg}; $\om_\eff$ in \cref{omeff};
the leading impact factors in \cref{phiTchoice,phiLchoice};
the NL ones in \cref{PhiT1om0} at $\om=0$ and \cref{phiT1om,phiL1om} at $\om\neq0$.

We shall compare our results with the experimental measurements of L3~\cite{L3:2001uuv} at $Q^2 = 16\ \text{GeV}^2$ and of OPAL~\cite{OPAL:2001fqu} at $Q^2 = 17.9\ \text{GeV}^2$, and also with previous 
calculations of the same cross section.
Since the values of $Q^2$ in L3 and OPAL are very close, it is reasonable to compare the data from both experiments with theoretical predictions at $Q^2 = 17\ \text{GeV}^2$.

We adopt the strong coupling value   to be 
$\alpha_s\left(Q^2=17 \ \text{GeV}^2\right) \approx 0.229$ as derived from the Particle Data Group \cite{Workman:2022ynf}.  

\begin{figure}[!htb]
\centering
\begin{minipage}[t]{0.48\textwidth}
    \centering
    \includegraphics[width=8.6cm]{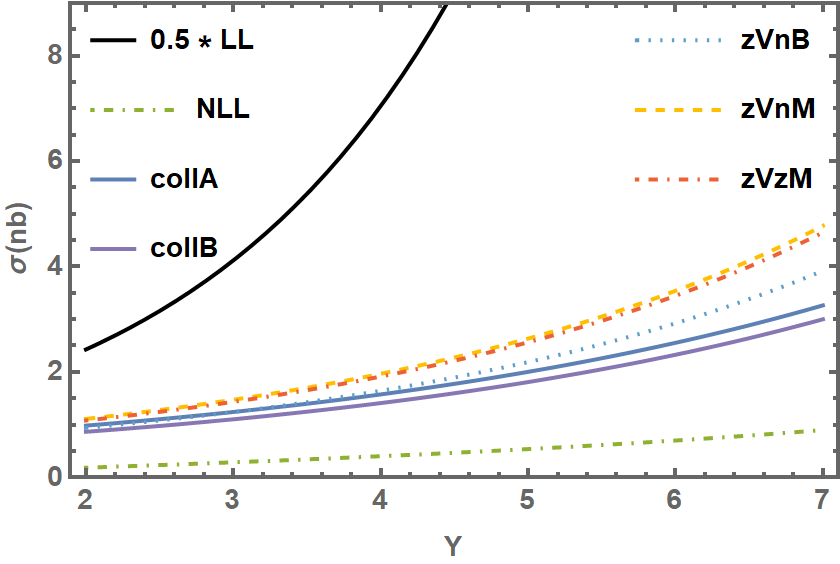}
    \caption{The value of the $\gamma^*\gamma^*$ cross section contribution from the BFKL exchange  for $Q^2 = 17\ \text{GeV}^2$ as a function of rapidity $Y$.  All five schemes (see \cref{t:Phi1Tschemes,t:Phi1Lschemes}) for the  NLL RGI calculation are shown together with the pure LL calculation (black solid and rescaled with a factor 0.5) and pure NLL calculation (green dot-dashed). }
    \label{fig: all 5 schemes}
\end{minipage}
\hfill
\begin{minipage}[t]{0.48\textwidth}
\centering
    \includegraphics[width=8.6cm]{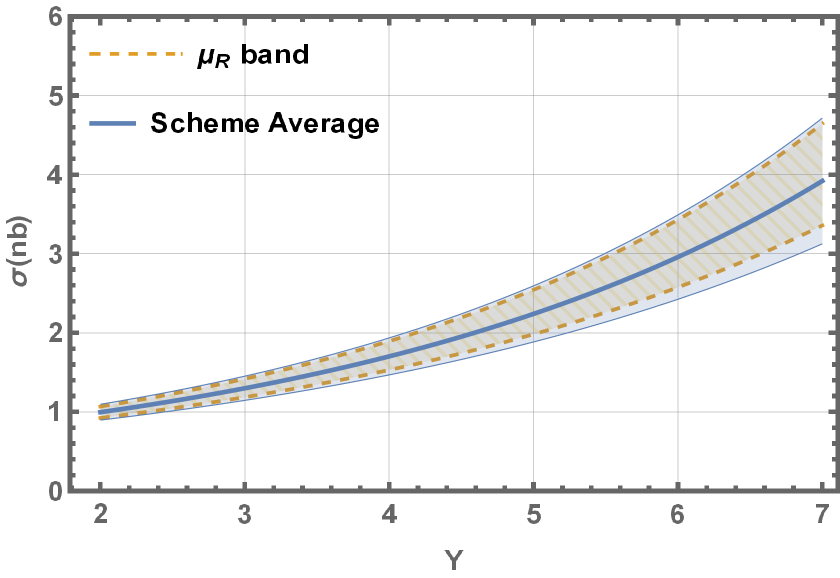}
    \caption{The value of the $\gamma^*\gamma^*$ cross section contribution from the BFKL exchange for $Q^2 = 17\ \text{GeV}^2$ as a function of rapidity $Y$. The scheme average band (blue-solid) represents the average value and standard deviation of the five resummed schemes. The $\mu_R$  band (yellow-dashed) is computed from average values of the five resummed schemes with half or double $\mu_R^2$ respectively.}
    \label{fig: scheme average band}
\end{minipage}
\end{figure}

In \cref{fig: all 5 schemes} we show the results for the NLL RGI cross sections using scheme I for the LO impact factors \cref{phiTchoice,phiLchoice} and the five different schemes from \cref{t:Phi1Tschemes,t:Phi1Lschemes} at NLO, and compare them with the pure LL and NLL cross sections. 

All five NLL  RGI cross sections are significantly reduced with respect to the LL calculation, however they are also significantly above the pure NLL calculation.
We observe that, the different schemes give very similar results.
 In order to present the results more intuitively, we incorporate a band to represent the scheme ambiguity as in \cref{fig: scheme average band}. The band size is defined as the standard deviation calculated from the five schemes at each rapidity $Y$. In the following, if the improved NLL cross section is presented as a single curve, then the curve is just the average for the five NLL RGI schemes. Adopting schemes II and III for the LO impact factors does not change significantly our estimates.

 In \cref{fig: scheme average band}, we also test the stability of the improved NLL cross section calculation with respect to the variation of the  $\mu_R$ scale. The upper and lower $\mu_R$ band is computed from average values of the five resummed schemes with half or double $\mu^2_R$ 
respectively. It turns out that the $\mu_R$ band size is slightly smaller than the scheme ambiguity band size. It is worth noting that besides the dependence on $\mu_R$ of the NLO impact factor and the running coupling argument, the NLO BFKL eigenfunction would also rely on $\mu_R$ when $\mu_R^2 \neq Q_1Q_2$,
\begin{equation}
    \tilde{X}_1(\om,\gamma) = X_1(\om,\gamma)+ \bbar\,
    X_0(\om,\gamma)\ln{\frac{\mu_R^2}{Q_1 Q_2}}.
\end{equation}
and the resummed effective $\omega$ after the NLO subtraction with $\mu_R$ dependency is then the solution of
\begin{equation}
    \omega = \asb(\mu_R^2) X_0(\omega,\gamma) + \asb^2(\mu_R^2)\left[ X_1(\omega,\gamma)+\bbar\, 
    X_0(\om,\gamma)\ln{\frac{\mu_R^2}{Q_1 Q_2}} \right].
\end{equation}

\begin{figure}[!htb]
\centering
\begin{minipage}[t]{0.48\textwidth}
    \centering
    \includegraphics[width=8.6cm]{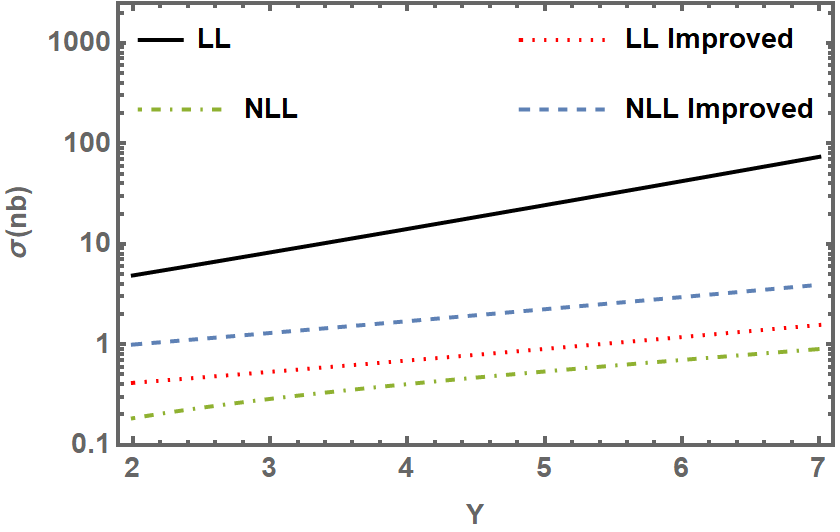}
    \caption{The value of the $\gamma^*\gamma^*$ cross section contribution from the BFKL exchange for $Q^2 = 17\ \text{GeV}^2$ as a function of rapidity in the logarithmic vertical scale. Pure LL is shown in black-solid,   NLL in green dashed-dotted, LL improved in red-dotted and NLL improved in blue-dashed. The NLL improved curve is the average of our five resummed NLL schemes (see text). }
    \label{fig: LO and NLO in log}
\end{minipage}
\hfill
\begin{minipage}[t]{0.48\textwidth}
\centering
    \centering
    \includegraphics[width=8.6cm]{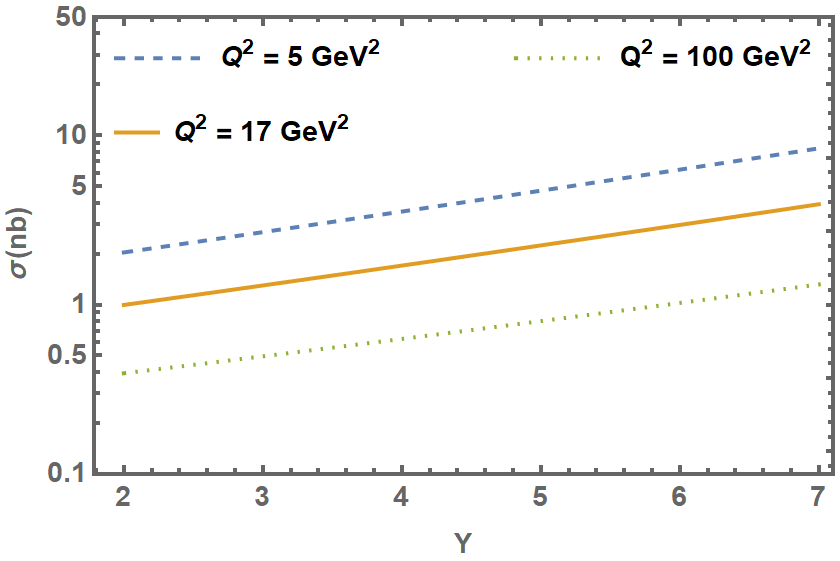}
    \caption{The value of the $\gamma^*\gamma^*$ cross section contribution from the BFKL exchange from NLL RGI calculation for $Q^2 = 5$ (blue-dashed),\ $17$ (yellow-solid), \ $100$\ $\text{GeV}^2$ (green-dotted) as a function of rapidity $Y$ in logarithmic vertical scale.  }
    \label{fig: different Q2}
\end{minipage}
\end{figure}

In \cref{fig: LO and NLO in log}, we compare the pure LL and NLL results (the latter computed using expressions from~refs.~\cite{Chirilli:2014dcb,chirilli2015erratum}), with the improved LL and NLL cross sections. Note the logarithmic vertical scale, which makes the characteristic exponential dependence of the cross section on the rapidity clearly visible. The NLL improved curve is given as the average of different schemes as explained above.

The improved LL and NLL calculations both tame the quick growth of the pure LL cross section with rapidity. It is worth noting that the improvement at LL alone --- consisting in the $\omega$ shifted LO eigenfunction and LO impact factors --- brings the curve down significantly.  We also observe that, the improved NLL is higher than the improved LL calculation, mostly because the improved NLO corrections bring a positive $O(\alpha_s^2)$ term to the impact factors. Finally we observe that improved calculations (both at LL and NLL) are above the pure NLL cross section.

In \cref{fig: different Q2}, we compare NLL RGI cross sections for $Q^2 = 5,\ 17, \ 100 \ \text{GeV}^2$. The cross section is strongly dependent on $Q^2$. The growth with rapidity is slowed down with increasing $Q^2$ due to the smaller value of the coupling constant, which affects the value of the leading exponent in the gluon Greens's function.

So far we have shown the contribution to the $\gamma^*\gamma^*$ cross section stemming only from the gluon exchanges, resummed by the BFKL evolution, which should be the dominant contribution at high energies. However, at lower energies, another contribution is important, namely the one from the `quark box' diagram. This contribution decreases with the rapidity, however it becomes dominant at low rapidities and is important when comparing with the experimental data. In the following, we evaluate the quark box in the lowest order \cite{Budnev:1975poe,Schienbein:2002wj}.  The total $\gamma^* \gamma^*$ cross section presented in the following includes both the quark box and the BFKL contributions.

\begin{figure}[!htb]
    \centering
    \includegraphics[width=0.8\linewidth]{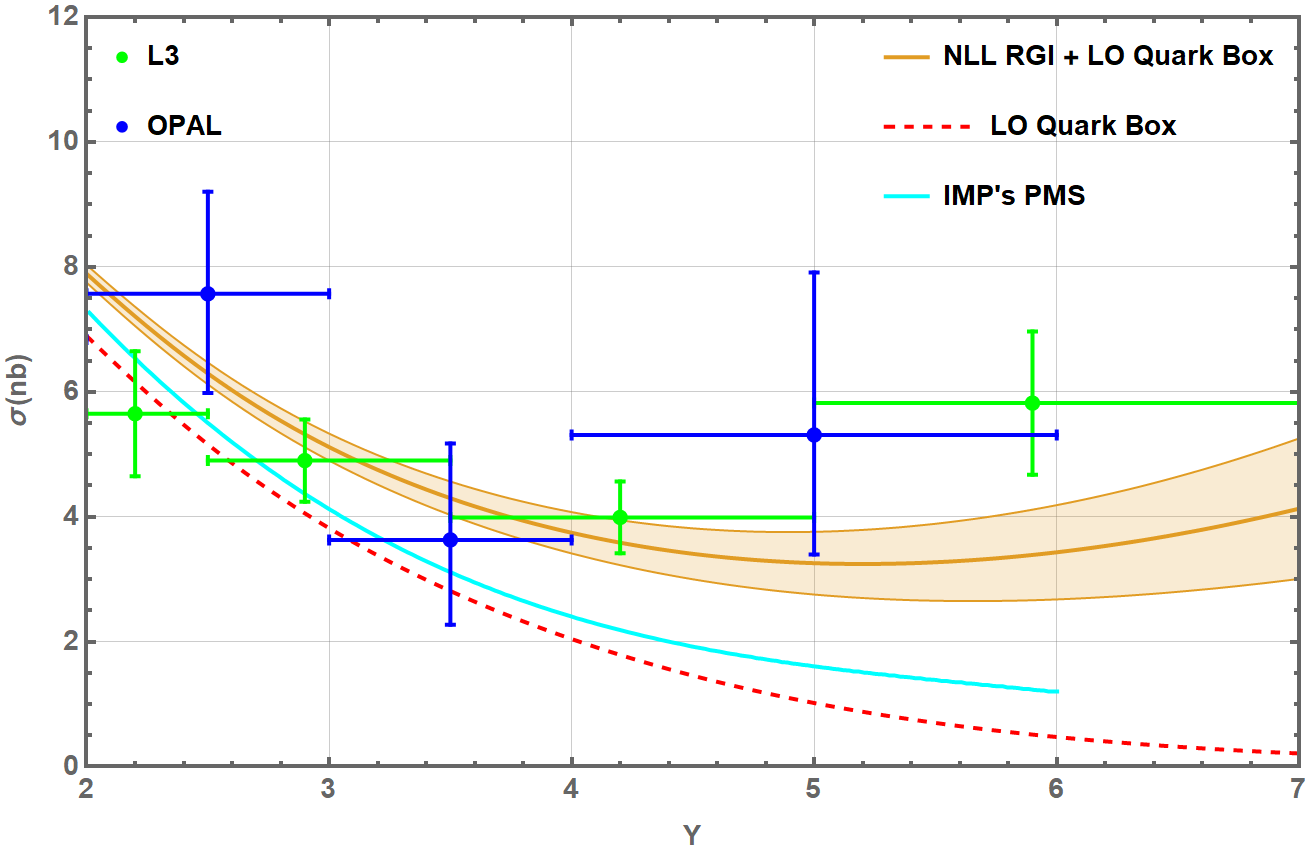}
    \caption{Cross sections for $Q^2 = 17\ \text{GeV}^2$, compared with L3 ($Q^2 = 16 \ \text{GeV}^2$) \cite{L3:2001uuv} and  OPAL ($Q^2 = 17.9\ \text{GeV}^2$) \cite{OPAL:2001fqu} data. The NLL improved curve is the sum of our averaged NLL BFKL resummed scheme and LO quark box contribution. The band size represents a combination of the scheme uncertainty and the $\mu_R$ band, i.e. $\delta_{\text{total}} = \sqrt{\delta_{\text{scheme}}^2+\delta_{\mu_R}^2}$. The calculation is done for $N_f=4$ massless flavours. The Ivanov-Murdaca-Papa's (IMP's) PMS optimized curve (solid-cyan) is from \cite{Ivanov:2014hpa}. Separately shown is the quark box contribution (dashed red).}
    \label{fig:Q17}
\end{figure}

In \cref{fig:Q17} 
we compare the results from NLL improved calculation with the experimental measurements of L3~\cite{L3:2001uuv} at $Q^2 = 16\ \text{GeV}^2$ and of OPAL~\cite{OPAL:2001fqu} at $Q^2 = 17.9\ \text{GeV}^2$, and also with previous 
calculations of the same cross section from \cite{Ivanov:2014hpa}.
As mentioned before, 
since the values of $Q^2$ in L3 and OPAL are very close, and the errors on the data points are such that  $Q^2$ dependence is not visible, it is reasonable to compare the data from both experiments with theoretical predictions at $Q^2 = 17\ \text{GeV}^2$.
We also show the LO quark box contribution in this figure.  We observe from \cref{fig:Q17} that the RGI NLL improved calculation has a stronger increase over rapidities than the pure NLL one. We also see that our result is significantly higher than the calculation from \cite{Ivanov:2014hpa}, particularly at high rapidities.
The RGI calculation is consistent with the experimental data from LEP  within the theoretical and experimental uncertainties. 

In the calculations we assumed $N_f=4$ and treated charm as massless. This is of course an approximation as the mass of the charm is expected to have some impact and to reduce the theoretical calculations. This effect was for example studied in detail in \cite{Bartels:2000sk}  and it was shown to decrease the cross section somewhat. However, this study was performed for the BFKL exchange at the leading order only. The full calculation would require small-$x$ resummation including the mass effects. This is an interesting problem in itself but it is beyond the scope of the current work.

\section{Conclusions\label{s:conclusions}}

In this paper, we have applied the collinear resummation, based on the
renormalization group improvement (RGI) for high-energy processes, to the
$\gamma^*\gamma^*$ cross section. The RGI formulation is based on a
factorization formula whose structure is similar to the one in the BFKL
approach, but whose impact factors explicitly depend on the Mellin variable
$\om$, conjugated to the center of mass energy squared $s$.
Firstly, we have computed RGI impact factors for transverse photons at LO and
NLO, which are consistent with the BFKL one in the high-energy limit and with
LO DGLAP in the collinear limit. We then extended this procedure for the
determination of the RGI impact factor for longitudinal photons.

At LO, the RGI impact factors are consistent with the impact factors with exact kinematics
computed in~\cite{Bialas:2001ks}. This is a non-trivial check, since we reproduce
the shifted position of the collinear poles in $\gamma$ --- the Mellin variable
conjugated to the photon virtualities --- and also the
coefficients of the leading $\gamma$-poles (apart from a subleading term
which is out of control in the approximations adopted in~\cite{Bialas:2001ks}).

At NLO we predict, and thus resum, the spurious energy-scale dependent quartic
(cubic) $\gamma$-poles of the transverse (longitudinal) BFKL impact factors.
For the transverse impact factors we can predict the cubic $\gamma$-poles, which
have both physical and spurious components. Having identified the physical
component of such poles, stemming from partonic anomalous dimensions and running of the coupling, we obtain
an impact factor which, in the collinear limit, is less singular than its BFKL
counterpart and contains the full LO DGLAP information.

In the case of the longitudinal photon impact factor the collinear analysis at NLO predicts a term proportional to the Casimir $C_F$ which is enhanced in the collinear region but is absent in the BFKL impact factors originally computed in \cite{Balitsky:2012bs} and presented in \cite{Chirilli:2014dcb,Ivanov:2014hpa}. This term stems from the $\ord{\as}$ coefficient function $C_{L,q}$ that couples the longitudinal photon to quarks at first order in QCD. We find full compatibility with the other colour structures, which quantitatively make up the bulk of the longitudinal impact factor.

Finally, we compute the $\gamma^*\gamma^*$ cross section using the NL RGI impact factors and Green's function, and compare it with previous calculations and also with experimental data.
Note that the RGI resummation suffers from some ambiguities, due to the lack of information in the kinematical region of low-energies (not controlled by BFKL) and comparable virtualities (not controlled by DGLAP) of the exchanged partons. Therefore, we propose a handful of physically motivated resummation schemes, and consider the average of the ensuing cross sections as our best estimate (with the corresponding standard deviation as resummation-scheme-uncertainty).

According to the expectations, the resummed cross section increases as a function of $Y=\log(s/Q_1 Q_2)$. It is found between the pure LL BFKL prediction and the pure NLL BFKL one. Note that switching from LL to NLL reduces the cross section by more than one order of magnitude. The resummation-scheme uncertainty is about 20\%, slightly larger than the renormalization-scheme uncertainty.

In order to provide a phenomenologically meaningful observable, the BFKL cross section, which is expected to dominate at large $Y$, must be supplemented with the so-called quark-box contribution, which dominates at small $Y$ and rapidly decreases with increasing $Y$. In this way, we are consistent with the experimental data of OPAL and L3, without the need of particular choices of running coupling scale fixing. This is a strong indication that the RGI procedure is the proper context for a correct description of virtual-photon scattering at large energies.

In the present calculation, we treat all four flavours, including charm, as massless. A more detailed analysis would need to include the charm mass in the full scheme of small $x$ resummation, which would require the knowledge of the massive impact factors at NLL. The mass effects at NLO in DIS structure functions were in fact computed in the dipole picture of high energy  \cite{Beuf:2021qqa,Beuf:2021srj,Beuf:2022ndu},  however this result would require extensive calculation (linearization to two gluon exchange as well as transforming it to momentum space) in order to extract the impact factors. 
A more detailed investigation on this issue is left for the future.

\section*{Acknowledgments}

We thank Ian Balitsky and Stefano Catani for the discussions.
This project is supported by the U.S. Department of Energy Grant DE-SC-0002145 and within the framework of the Saturated Glue (SURGE) Topical Theory Collaboration,  by Polish NCN  Grant No. 2019/33/B/ST2/02588, and by
 the European Union’s Horizon
  2020 research and innovation programme under grant agreement No
  824093.

\appendix

\section{Lowest-order cross sections and structure functions\label{s:locs}}

In this appendix, we sketch the determination of the photon-parton cross sections at
the lowest order in perturbation theory, which is the basis of the analysis of the
photon-photon cross section in the collinear regime $Q_1^2\ll Q_2^2$.
Such cross sections are proportional to the corresponding partonic structure functions,
which in turn can be derived by the DIS coefficient functions.

\subsection{Transverse photon\label{s:trph}}

The cross section of a virtual photon with polarization $\lambda$ scattering on
a particle of momentum $P$ (e.g., a hadron) is given by (cfr.~\cite{Bialas:2001ks})
\begin{align}
  \sigma^{(\lambda)}(P,q) = \frac{4\pi^2\alpha}{Q^2} F^{(\lambda)}(x,Q^2) \;,
  \qquad x\dug\frac{Q^2}{2P\cdot q} \;.
\end{align}
where $F^{(\lambda)}(x,Q^2):\lambda=L,T,2$ are the standard structure functions
with $F^{(2)}=F^{(L)}+F^{(T)}$.  The integrand $\dbtilde{\sigma}(\om,\gamma;p)$ of
the double Mellin representation~\eqref{sigmaRGI} can then be written as
\begin{align}
  \dbtilde{\sigma}(\om,\gamma;1) &= \int_{Q_1^2}^\infty\frac{\dif s}{s}
  \left(\frac{Q_1^2}{s}\right)^\om \int\frac{\dif Q_1^2}{Q_1^2}
  \left(\frac{Q_2^2}{Q_1^2}\right)^{\gamma-\half} 2\pi Q_1 Q_2
  \,\sigma(s,Q_1^2,Q_2^2) \nonumber \\
  &= \int_0^1\frac{\dif x}{x} x^{\om}
  \int\frac{\dif Q_1^2}{Q_1^2} \left(\frac{Q_2^2}{Q_1^2}\right)^{\gamma}
  (2\pi)^3 \alpha F(x,Q_1^2) \;. \label{sigmaF}
\end{align}

In the case of an incoming quark of flavour $a$ and small off\-shell\-ness
$Q_2^2\ll Q_1^2\equiv Q^2$, the partonic structure functions at lowest order are
nothing but the corresponding coefficient functions:
\begin{equation}
  F^{(T,a)}_0(x,Q^2) = F^{(2,a)}_0(x,Q^2) = x\; e_a^2 \; C^{(2,q)}_0(x)
  = e_a^2 \delta(1-x) \;,\qquad
  F^{(L,a)}_0(x,Q^2) = 0 \;.
\end{equation}
Therefore, at the lowest order only the transverse polarization is effective and we have
\begin{equation}\label{IT0}
  \dbtilde{\sigma}^{(T,a)}_0(\om,\gamma;1)
  = \int_0^1\frac{\dif x}{x} x^{\om}
  \int\frac{\dif Q_1^2}{Q_1^2} \left(\frac{Q_2^2}{Q_1^2}\right)^{\gamma}
  (2\pi)^3 \alpha F^{(T,a)}_0(x,Q_1^2)
  =(2\pi)^3 \alpha\, e_a^2 \frac1{\gamma}
\end{equation}
which is the first factor of the collinear
chain~\eqref{colChainLO}, before summing over quark and antiquark flavours.
By taking the inverse Mellin transform with respect to $\gamma$, we have
\begin{equation}\label{sigT0a}
  \dbtilde{\sigma}^{(T,a)}_0(\om,Q_1^2,Q_2^2;1)
  =(2\pi)^3 \alpha\, e_a^2 \;,
\end{equation}
representing the first factor in \cref{ITTQ} --- again, before summing over quark and antiquark flavours.

The first non-vanishing contribution of the photon-gluon structure functions
starts at $\ord{\as}$. In the collinear limit, i.e., considering the strong ordering
of partons' momenta, each rung provides a factor
$\int_{k_{i-1}^2}^{k_{i}^2}\frac{\dif k^2}{k^2} \frac{\as(k^2)}{2\pi} P_{ab}(\om)$. With fixed running coupling such a factor reduces to
$\frac{\as}{2\pi} \log\frac{k_i^2}{k_{i-1}^2} P_{ab}(\om)$, which
becomes $\frac{\as}{2\pi} P_{ab}(\om)/\gamma$ in $\gamma$-space.
Therefore, at $\ord{\as}$, for a transverse photon we have
\begin{equation}\label{ITg1}
  \dbtilde{\sigma}_1^{(T,g)}(\om,\gamma;1) = \sum_a (2\pi)^3 \alpha\, e_a^2
  \frac1{\gamma} \cdot \frac{\as}{2\pi} \frac{P_{(q=a)g}(\om)}{\gamma} \;.
\end{equation}
in agreement with the first factors of \cref{colChainLO}
since $\sum_a e_a^2 = 2\sum_q e_q^2$. The other factors follow from the remaining two vertices.

\subsection{Longitudinal photon\label{s:loph}}

The longitudinal structure function starts at $\ord{\alpha\as}$ in perturbation
theory, and receive contributions from gluons and quarks. We start by
considering the gluon-initiated structure function, which is well known in the
literature, and can be read, e.g., from eq.~(B.5) of ref.~\cite{Zijlstra:1992qd}
\begin{align}
  F_1^{(L,g)}(x,Q^2) = x\; \frac{\sum_a e_a^2}{2N_f}\;C_1^{(L,g)}(x,Q^2/\mu_F^2)
  =\frac{\as}{2\pi} \big(\sum_a e_a^2\big)\, T_R\, 4\,x^2(1-x) \;.
\end{align}
The corresponding longitudinal photon-gluon cross section in Mellin space can
then be determined from \cref{sigmaF} and reads
\begin{equation}\label{ILg1}
  \dbtilde\sigma^{(L,g)}_1(\om,\gamma;1) =
  \frac{16\pi^2 \alpha\as\,\big(\sum_a e_a^2\big)\,T_R}{\gamma\,(2+\om)(3+\om)} \;.
\end{equation}
It is straightforward to check that the r.h.s.\ of \cref{ILg1} is
proportional to the simple pole at $\gamma=0$ of the BNP longitudinal
impact factor with exact kinematics with its full $\om$-dependence,
just like the r.h.s.\ of \cref{ITg1} is proportional to the
double pole of the BNP transverse impact factor:
\begin{align}
  \dbtilde\sigma^{(T,g)}_1(\om,\gamma;1) &= 2\pi\big(\sum_a e_a^2\big)T_R
  \,S_T(\om,\gamma) + \ord{\gamma^{-1}} \\
  \dbtilde\sigma^{(L,g)}_1(\om,\gamma;1) &= 2\pi\big(\sum_a e_a^2\big)T_R
  \,S_L(\om,\gamma) + \ord{\gamma^0} \;.
\end{align}

The quark-initiated structure function, is also well known in the
literature, and can be read, e.g., from eq.~(B.1) of ref.~\cite{Zijlstra:1992qd}:
\begin{align}
  F_1^{(L,a)}(x,Q^2) = x\; e_a^2\;C_1^{(L,q)}(x,Q^2/\mu_F^2)
  =\frac{\as}{2\pi} e_a^2\, C_F\,2 x^2
  \qquad \text{($a =$ quark or antiquark)}
  \;.
\end{align}
The corresponding longitudinal photon-quark cross section in Mellin space can
then be determined from \cref{sigmaF} and reads
\begin{equation}\label{ILq1}
  \dbtilde\sigma^{(L,a)}_1(\om,\gamma;1) =
  \frac{8\pi^2 \alpha\as\, e_a^2\,C_F}{\gamma\,(2+\om)} 
  \qquad \text{($a =$ quark or antiquark)}\;.
\end{equation}
Summing over all quarks and antiquarks we get
\begin{equation}\label{LqLgFact}
  \sum_a \dbtilde\sigma^{(L,a)}_1(\om,\gamma;1) = \frac{C_F}{T_R}\,\frac{3+\om}{2} \,\dbtilde\sigma^{(L,g)}_1(\om,\gamma;1) \;. 
\end{equation}
In practice, the blob connecting a longitudinal photon to all quarks and antiquarks displayed in  \cref{f:longColChainNLO}(c) is equal to the blob connecting the longitudinal photon to a gluon in \cref{f:longColChainNLO}(a),(b) up to the additional multiplicative factor $C_F (3+\om)/(2 T_R)$.

\bibliography{mybib}

\begin{thebibliography}{10}

\bibitem{Catani:1990eg}
S.~Catani, M.~Ciafaloni, and F.~Hautmann, ``{High-energy factorization and
  small x heavy flavor production},'' {\em Nucl. Phys. B}, vol.~366,
  pp.~135--188, 1991.

\bibitem{Catani:1990xk}
S.~Catani, M.~Ciafaloni, and F.~Hautmann, ``{Gluon contributions to small x
  heavy flavor production},'' {\em Phys. Lett. B}, vol.~242, pp.~97--102, 1990.

\bibitem{Catani:1994sq}
S.~Catani and F.~Hautmann, ``{High-energy factorization and small x deep
  inelastic scattering beyond leading order},'' {\em Nucl. Phys. B}, vol.~427,
  pp.~475--524, 1994.

\bibitem{Kuraev:1977fs}
E.~A. Kuraev, L.~N. Lipatov, and R.~S. Fadin, ``{The Pomeranchuk Singularity in
  Nonabelian Gauge Theories},'' {\em Sov. Phys. JETP}, vol.~45, pp.~199--204,
  1977.
\newblock [Zh. Eksp. Teor. Fiz.72,377(1977)].

\bibitem{Balitsky:1978ic}
I.~I. Balitsky and L.~N. Lipatov, ``{The Pomeranchuk Singularity in Quantum
  Chromodynamics},'' {\em Sov. J. Nucl. Phys.}, vol.~28, pp.~822--829, 1978.
\newblock [Yad. Fiz.28,1597(1978)].

\bibitem{Lipatov:1985uk}
L.~N. Lipatov, ``{The Bare Pomeron in Quantum Chromodynamics},'' {\em Sov.
  Phys. JETP}, vol.~63, pp.~904--912, 1986.
\newblock [Zh. Eksp. Teor. Fiz.90,1536(1986)].

\bibitem{Fadin:1998py}
V.~S. Fadin and L.~N. Lipatov, ``{BFKL pomeron in the next-to-leading
  approximation},'' {\em Phys. Lett. B}, vol.~429, pp.~127--134, 1998.

\bibitem{Ciafaloni:1998gs}
M.~Ciafaloni and G.~Camici, ``{Energy scale(s) and next-to-leading BFKL
  equation},'' {\em Phys. Lett. B}, vol.~430, pp.~349--354, 1998.

\bibitem{Salam:1998tj}
G.~P. Salam, ``{A Resummation of large subleading corrections at small x},''
  {\em JHEP}, vol.~07, p.~019, 1998.

\bibitem{Salam:1999cn}
G.~P. Salam, ``{An Introduction to leading and next-to-leading BFKL},'' {\em
  Acta Phys. Polon.}, vol.~B30, pp.~3679--3705, 1999.

\bibitem{Altarelli:1999vw}
G.~Altarelli, R.~D. Ball, and S.~Forte, ``{Resummation of singlet parton
  evolution at small x},'' {\em Nucl. Phys.}, vol.~B575, pp.~313--329, 2000.

\bibitem{Altarelli:2000mh}
G.~Altarelli, R.~D. Ball, and S.~Forte, ``{Small x resummation and HERA
  structure function data},'' {\em Nucl. Phys.}, vol.~B599, pp.~383--423, 2001.

\bibitem{Altarelli:2001ji}
G.~Altarelli, R.~D. Ball, and S.~Forte, ``{Factorization and resummation of
  small x scaling violations with running coupling},'' {\em Nucl. Phys.},
  vol.~B621, pp.~359--387, 2002.

\bibitem{Altarelli:2003hk}
G.~Altarelli, R.~D. Ball, and S.~Forte, ``{An Anomalous dimension for small x
  evolution},'' {\em Nucl. Phys.}, vol.~B674, pp.~459--483, 2003.

\bibitem{Altarelli:2008aj}
G.~Altarelli, R.~D. Ball, and S.~Forte, ``{Small x Resummation with Quarks:
  Deep-Inelastic Scattering},'' {\em Nucl. Phys.}, vol.~B799, pp.~199--240,
  2008.

\bibitem{Ciafaloni:1999au}
M.~Ciafaloni, D.~Colferai, and G.~P. Salam, ``{A collinear model for small x
  physics},'' {\em JHEP}, vol.~10, p.~017, 1999.

\bibitem{Ciafaloni:1999yw}
M.~Ciafaloni, D.~Colferai, and G.~P. Salam, ``{Renormalization group improved
  small x equation},'' {\em Phys. Rev.}, vol.~D60, p.~114036, 1999.

\bibitem{Ciafaloni:2003ek}
M.~Ciafaloni, D.~Colferai, D.~Colferai, G.~P. Salam, and A.~M. Stasto,
  ``{Extending QCD perturbation theory to higher energies},'' {\em Phys. Lett.
  B}, vol.~576, pp.~143--151, 2003.

\bibitem{Ciafaloni:2003rd}
M.~Ciafaloni, D.~Colferai, G.~P. Salam, and A.~M. Stasto, ``{Renormalization
  group improved small x Green's function},'' {\em Phys. Rev. D}, vol.~68,
  p.~114003, 2003.

\bibitem{Ciafaloni:2003kd}
M.~Ciafaloni, D.~Colferai, G.~P. Salam, and A.~M. Stasto, ``{The Gluon
  splitting function at moderately small x},'' {\em Phys. Lett.}, vol.~B587,
  pp.~87--94, 2004.

\bibitem{Ciafaloni:2007gf}
M.~Ciafaloni, D.~Colferai, G.~P. Salam, and A.~M. Stasto, ``{A Matrix
  formulation for small-x singlet evolution},'' {\em JHEP}, vol.~08, p.~046,
  2007.

\bibitem{Thorne:2001nr}
R.~S. Thorne, ``{The Running coupling BFKL anomalous dimensions and splitting
  functions},'' {\em Phys. Rev.}, vol.~D64, p.~074005, 2001.

\bibitem{SabioVera:2005tiv}
A.~Sabio~Vera, ``{An 'All-poles' approximation to collinear resummations in the
  Regge limit of perturbative QCD},'' {\em Nucl. Phys. B}, vol.~722,
  pp.~65--80, 2005.

\bibitem{Bonvini:2016wki}
M.~Bonvini, S.~Marzani, and T.~Peraro, ``{Small-$x$ resummation from HELL},''
  {\em Eur. Phys. J. C}, vol.~76, no.~11, p.~597, 2016.

\bibitem{Bonvini:2017ogt}
M.~Bonvini, S.~Marzani, and C.~Muselli, ``{Towards parton distribution
  functions with small-$x$ resummation: HELL 2.0},'' {\em JHEP}, vol.~12,
  p.~117, 2017.

\bibitem{Ball:2017otu}
R.~D. Ball, V.~Bertone, M.~Bonvini, S.~Marzani, J.~Rojo, and L.~Rottoli,
  ``{Parton distributions with small-x resummation: evidence for BFKL dynamics
  in HERA data},'' {\em Eur. Phys. J.}, vol.~C78, no.~4, p.~321, 2018.

\bibitem{Kwiecinski:1996td}
J.~Kwiecinski, A.~D. Martin, and P.~J. Sutton, ``{Constraints on gluon
  evolution at small x},'' {\em Z. Phys.}, vol.~C71, pp.~585--594, 1996.

\bibitem{Bartels:2002uz}
J.~Bartels, D.~Colferai, S.~Gieseke, and A.~Kyrieleis, ``{NLO corrections to
  the photon impact factor: Combining real and virtual corrections},'' {\em
  Phys. Rev. D}, vol.~66, p.~094017, 2002.

\bibitem{Balitsky:2012bs}
I.~Balitsky and G.~A. Chirilli, ``{Photon impact factor and $k_T$-factorization
  for DIS in the next-to-leading order},'' {\em Phys. Rev. D}, vol.~87, no.~1,
  p.~014013, 2013.

\bibitem{Colferai:2010wu}
D.~Colferai, F.~Schwennsen, L.~Szymanowski, and S.~Wallon, ``{Mueller Navelet
  jets at LHC - complete NLL BFKL calculation},'' {\em JHEP}, vol.~12, p.~026,
  2010.

\bibitem{Hentschinski:2014bra}
M.~Hentschinski, J.~D. Madrigal~Mart\'\i{}nez, B.~Murdaca, and A.~Sabio~Vera,
  ``{The quark induced Mueller\textendash{}Tang jet impact factor at
  next-to-leading order},'' {\em Nucl. Phys. B}, vol.~887, pp.~309--337, 2014.

\bibitem{Hentschinski:2014esa}
M.~Hentschinski, J.~D.~M. Mart\'\i{}nez, B.~Murdaca, and A.~Sabio~Vera, ``{The
  gluon-induced Mueller\textendash{}Tang jet impact factor at next-to-leading
  order},'' {\em Nucl. Phys. B}, vol.~889, pp.~549--579, 2014.

\bibitem{Colferai:2023hje}
D.~Colferai, F.~Deganutti, T.~G. Raben, and C.~Royon, ``{First computation of
  Mueller Tang processes using a full NLL BFKL approach},'' {\em JHEP},
  vol.~06, p.~091, 2023.

\bibitem{Ivanov:2004pp}
D.~Y. Ivanov, M.~I. Kotsky, and A.~Papa, ``{The Impact factor for the virtual
  photon to light vector meson transition},'' {\em Eur. Phys. J. C}, vol.~38,
  pp.~195--213, 2004.

\bibitem{Beuf:2016wdz}
G.~Beuf, ``{Dipole factorization for DIS at NLO: Loop correction to the
  $\gamma^*_{T,L}\to q\overline q$ light-front wave functions},'' {\em Phys.
  Rev. D}, vol.~94, no.~5, p.~054016, 2016.

\bibitem{Beuf:2017bpd}
G.~Beuf, ``{Dipole factorization for DIS at NLO: Combining the $q\bar{q}$ and
  $q\bar{q}g$ contributions},'' {\em Phys. Rev. D}, vol.~96, no.~7, p.~074033,
  2017.

\bibitem{Beuf:2022ndu}
G.~Beuf, T.~Lappi, and R.~Paatelainen, ``{Massive quarks in NLO dipole
  factorization for DIS: Transverse photon},'' {\em Phys. Rev. D}, vol.~106,
  no.~3, p.~034013, 2022.

\bibitem{Beuf:2022kyp}
G.~Beuf, H.~H\"anninen, T.~Lappi, Y.~Mulian, and H.~M\"antysaari,
  ``{Diffractive deep inelastic scattering at NLO in the dipole picture: The
  $q\bar{q}g$ contribution},'' {\em Phys. Rev. D}, vol.~106, no.~9, p.~094014,
  2022.

\bibitem{Boussarie:2016bkq}
R.~Boussarie, A.~V. Grabovsky, D.~Y. Ivanov, L.~Szymanowski, and S.~Wallon,
  ``{Next-to-Leading Order Computation of Exclusive Diffractive Light Vector
  Meson Production in a Saturation Framework},'' {\em Phys. Rev. Lett.},
  vol.~119, no.~7, p.~072002, 2017.

\bibitem{Boussarie:2019ero}
R.~Boussarie, A.~V. Grabovsky, L.~Szymanowski, and S.~Wallon, ``{Towards a
  complete next-to-logarithmic description of forward exclusive diffractive
  dijet electroproduction at HERA: real corrections},'' {\em Phys. Rev. D},
  vol.~100, no.~7, p.~074020, 2019.

\bibitem{Mantysaari:2022kdm}
H.~M\"antysaari and J.~Penttala, ``{Complete calculation of exclusive heavy
  vector meson production at next-to-leading order in the dipole picture},''
  {\em JHEP}, vol.~08, p.~247, 2022.

\bibitem{Mantysaari:2022bsp}
H.~M\"antysaari and J.~Penttala, ``{Exclusive production of light vector mesons
  at next-to-leading order in the dipole picture},'' {\em Phys. Rev. D},
  vol.~105, no.~11, p.~114038, 2022.

\bibitem{Caucal:2022ulg}
P.~Caucal, F.~Salazar, B.~Schenke, and R.~Venugopalan, ``{Back-to-back
  inclusive dijets in DIS at small x: Sudakov suppression and gluon saturation
  at NLO},'' {\em JHEP}, vol.~11, p.~169, 2022.

\bibitem{Roy:2019hwr}
K.~Roy and R.~Venugopalan, ``{NLO impact factor for inclusive photon$+$dijet
  production in $e+A$ DIS at small $x$},'' {\em Phys. Rev. D}, vol.~101, no.~3,
  p.~034028, 2020.

\bibitem{Iancu:2015joa}
E.~Iancu, J.~D. Madrigal, A.~H. Mueller, G.~Soyez, and D.~N.
  Triantafyllopoulos, ``{Collinearly-improved BK evolution meets the HERA
  data},'' {\em Phys. Lett. B}, vol.~750, pp.~643--652, 2015.

\bibitem{Brodsky:1996sg}
S.~J. Brodsky, F.~Hautmann, and D.~E. Soper, ``{Probing the QCD pomeron in $e^+
  e^-$ collisions},'' {\em Phys. Rev. Lett.}, vol.~78, pp.~803--806, 1997.
\newblock [Erratum: Phys.Rev.Lett. 79, 3544 (1997)].

\bibitem{Brodsky:1997sd}
S.~J. Brodsky, F.~Hautmann, and D.~E. Soper, ``{Virtual photon scattering at
  high-energies as a probe of the short distance pomeron},'' {\em Phys. Rev.
  D}, vol.~56, pp.~6957--6979, 1997.

\bibitem{Bartels:1996ke}
J.~Bartels, A.~De~Roeck, and H.~Lotter, ``{The $\gamma^* \gamma^*$ total
  cross-section and the BFKL pomeron at $e^+ e^-$ colliders},'' {\em Phys.
  Lett. B}, vol.~389, pp.~742--748, 1996.

\bibitem{Bartels:2000sk}
J.~Bartels, C.~Ewerz, and R.~Staritzbichler, ``{Effect of the charm quark mass
  on the BFKL $\gamma^* \gamma^*$ total cross-section at LEP},'' {\em Phys.
  Lett. B}, vol.~492, pp.~56--62, 2000.

\bibitem{Donnachie:1999kp}
A.~Donnachie, H.~G. Dosch, and M.~Rueter, ``{$\gamma^* \gamma^*$ reactions at
  high-energies},'' {\em Eur. Phys. J. C}, vol.~13, pp.~141--150, 2000.

\bibitem{Donnachie:1999py}
A.~Donnachie and S.~Soldner-Rembold, ``{$\gamma^* \gamma^*$ reaction at
  high-energies},'' {\em J. Phys. G}, vol.~26, pp.~689--695, 2000.

\bibitem{Kwiecinski:1999yx}
J.~Kwiecinski and L.~Motyka, ``{Probing the QCD pomeron in doubly tagged $e^+
  e^-$ collisions},'' {\em Phys. Lett. B}, vol.~462, pp.~203--210, 1999.

\bibitem{Kwiecinski:2000zs}
J.~Kwiecinski and L.~Motyka, ``{Theoretical description of the total $\gamma^*
  \gamma^*$ cross-section and its confrontation with the LEP data on doubly
  tagged $e^+ e^-$ events},'' {\em Eur. Phys. J. C}, vol.~18, pp.~343--351,
  2000.

\bibitem{L3:2001uuv}
P.~Achard {\em et~al.}, ``{Double tag events in two photon collisions at
  LEP},'' {\em Phys. Lett. B}, vol.~531, pp.~39--51, 2002.

\bibitem{OPAL:2001fqu}
G.~Abbiendi {\em et~al.}, ``{Measurement of the hadronic cross-section for the
  scattering of two virtual photons at LEP},'' {\em Eur. Phys. J. C}, vol.~24,
  pp.~17--31, 2002.

\bibitem{Chirilli:2014dcb}
G.~A. Chirilli and Y.~V. Kovchegov, ``{$\gamma^* \gamma^*$ Cross Section at NLO
  and Properties of the BFKL Evolution at Higher Orders},'' {\em JHEP},
  vol.~05, p.~099, 2014.
\newblock [Erratum: JHEP 08, 075 (2015)].

\bibitem{Ivanov:2014hpa}
D.~Y. Ivanov, B.~Murdaca, and A.~Papa, ``{The $\gamma^* \gamma^*$ total cross
  section in next-to-leading order BFKL and LEP2 data},'' {\em JHEP}, vol.~10,
  p.~058, 2014.

\bibitem{Askew:1992tw}
A.~J. Askew, J.~Kwiecinski, A.~D. Martin, and P.~J. Sutton, ``{QCD predictions
  for deep inelastic structure functions at HERA},'' {\em Phys. Rev. D},
  vol.~47, pp.~3775--3782, 1993.

\bibitem{Askew:1993jk}
A.~J. Askew, J.~Kwiecinski, A.~D. Martin, and P.~J. Sutton, ``{Properties of
  the BFKL equation and structure function predictions for HERA},'' {\em Phys.
  Rev. D}, vol.~49, pp.~4402--4414, 1994.

\bibitem{Kwiecinski:1997ee}
J.~Kwiecinski, A.~D. Martin, and A.~M. Stasto, ``{A Unified BFKL and GLAP
  description of $F_2$ data},'' {\em Phys. Rev.}, vol.~D56, pp.~3991--4006,
  1997.

\bibitem{Bialas:2001ks}
A.~Bialas, H.~Navelet, and R.~B. Peschanski, ``{Virtual photon impact factors
  with exact gluon kinematics},'' {\em Nucl. Phys. B}, vol.~603, pp.~218--230,
  2001.

\bibitem{Velizhanin:2015xsa}
V.~N. Velizhanin, ``{BFKL pomeron in the next-to-next-to-leading approximation
  in the planar N=4 SYM theory},'' 2015.

\bibitem{Gromov:2015vua}
N.~Gromov, F.~Levkovich-Maslyuk, and G.~Sizov, ``{Pomeron Eigenvalue at Three
  Loops in $\mathcal N=$ 4 Supersymmetric Yang-Mills Theory},'' {\em Phys. Rev.
  Lett.}, vol.~115, no.~25, p.~251601, 2015.

\bibitem{caron2018high}
S.~Caron-Huot and M.~Herranen, ``High-energy evolution to three loops,'' {\em
  Journal of High Energy Physics}, vol.~2018, no.~2, p.~58, 2018.

\bibitem{Ciafaloni:1998iv}
M.~Ciafaloni and D.~Colferai, ``{The BFKL equation at next-to-leading level and
  beyond},'' {\em Phys. Lett. B}, vol.~452, pp.~372--378, 1999.

\bibitem{Deak:2019wms}
M.~Deak, K.~Kutak, W.~Li, and A.~M. Stasto, ``{On the different forms of the
  kinematical constraint in BFKL},'' {\em Eur. Phys. J.}, vol.~C79, no.~8,
  p.~647, 2019.

\bibitem{Chirilli:2013kca}
G.~A. Chirilli and Y.~V. Kovchegov, ``{Solution of the NLO BFKL Equation and a
  Strategy for Solving the All-Order BFKL Equation},'' {\em JHEP}, vol.~06,
  p.~055, 2013.

\bibitem{Workman:2022ynf}
R.~L. Workman and Others, ``{Review of Particle Physics},'' {\em PTEP},
  vol.~2022, p.~083C01, 2022.

\bibitem{chirilli2015erratum}
G.~Chirilli and Y.~Kovchegov, ``Erratum to: $\gamma^* \gamma^*$ cross section
  at nlo and properties of the bfkl evolution at higher orders,'' {\em Journal
  of High Energy Physics}, 2015.

\bibitem{Budnev:1975poe}
V.~M. Budnev, I.~F. Ginzburg, G.~V. Meledin, and V.~G. Serbo, ``{The Two photon
  particle production mechanism. Physical problems. Applications. Equivalent
  photon approximation},'' {\em Phys. Rept.}, vol.~15, pp.~181--281, 1975.

\bibitem{Schienbein:2002wj}
I.~Schienbein, ``{Two photon processes and photon structure},'' {\em Annals
  Phys.}, vol.~301, pp.~128--156, 2002.

\bibitem{Beuf:2021qqa}
G.~Beuf, T.~Lappi, and R.~Paatelainen, ``{Massive quarks in NLO dipole
  factorization for DIS: Longitudinal photon},'' {\em Phys. Rev. D}, vol.~104,
  no.~5, p.~056032, 2021.

\bibitem{Beuf:2021srj}
G.~Beuf, T.~Lappi, and R.~Paatelainen, ``{Massive Quarks at One Loop in the
  Dipole Picture of Deep Inelastic Scattering},'' {\em Phys. Rev. Lett.},
  vol.~129, no.~7, p.~072001, 2022.

\bibitem{Zijlstra:1992qd}
E.~B. Zijlstra and W.~L. van Neerven, ``{Order $\alpha_s^2$ QCD corrections to
  the deep inelastic proton structure functions $F_2$ and $F_L$},'' {\em Nucl.
  Phys. B}, vol.~383, pp.~525--574, 1992.

\end{thebibliography}
\bibliographystyle{ieeetr}

\end{document}